\documentclass[prb,preprint,showkeys,showpacs]{revtex4}

\usepackage{epsfig}

\usepackage{dcolumn}

\begin{document}

\title{Melting of icosahedral gold nanoclusters from molecular
       dynamics simulations}

\author{Yanting Wang and S. Teitel}

\affiliation{Department of Physics and Astronomy, University of
Rochester, Rochester, NY 14627}

\author{Christoph Dellago}

\affiliation{Institute for Experimental Physics, University of
Vienna, Boltzmanngasse 5, 1090 Vienna, Austria}

\date{\today}

\vspace*{1cm}
\begin{abstract}

Molecular dynamics simulations show that gold clusters with about 
$600$--$3000$ atoms crystallize into a Mackay icosahedron upon 
cooling from the liquid. A detailed surface analysis shows that the 
facets on the surface of the Mackay icosahedral gold clusters soften but do 
not premelt below the bulk melting temperature. 
This softening is found to be due 
to the increasing mobility of vertex and edge atoms with temperature,
which leads to inter-layer and intra-layer diffusion, and a shrinkage of the
average facet size, so that the average shape of the cluster is nearly
spherical at melting.

\end{abstract}

\pacs{}

\keywords{}

\maketitle


\section{INTRODUCTION}

Nanocrystals have quite different physical properties from their corresponding
bulk materials mainly because of their large surface-to-volume ratio. 
Among nobel metals, gold nanoparticles have already shown their 
promise for a wide range of applications, such as 
nano-lithography \cite{ZHENG}, catalysts \cite{BELL},
nano-bioelectronic devices \cite{WILLNER},
and ion detection \cite{MURPHY}. 
Thus knowledge of the structure and stability of gold nanocrystals 
is of great importance.

\begin{figure}
\epsfxsize=8.6truecm
\epsfbox{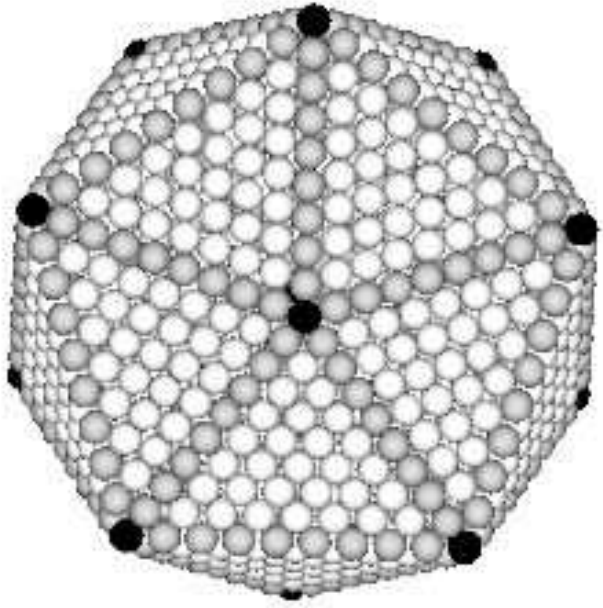}
\caption{An ideal Ih structure with the magic number
of $2869$ atoms. Atoms are shaded to indicate their local structure:
fcc is white, hcp is gray, and other is black.}
\label{fig:2869}
\end{figure}

While bulk gold has an fcc crystal structure, the competition between
bulk and surface energies in nanometer sized gold crystallites can result 
in several different competing structures \cite{IIJIMA,LANDMAN}. 
One such structure which 
has been observed both in simulations \cite{BARTELL,NAM} 
and in experiments \cite{MARKS_REVIEW,ASCENCIO} is
the ``Mackay icosahedron''\cite{MACKAY,SHELLS}, which we will also denote
as the ``Ih structure'',
consisting of $20$ slightly distorted
fcc tetrahedra, with four $\{111\}$ facets each, meeting at the 
center to form an icosahedral shaped cluster. The
internal facets of the tetrahedra meet at strain inducing
twin grain boundaries with a local hcp structure, 
leaving the cluster with $20$ external $\{111\}$
facets. For an Ih structure 
with $L$ shells, the magic number of atoms needed to construct a perfectly
symmetric ideal icosahedron is \cite{SHELLS}, 
\begin{equation}
  N_L = \frac{10}{3}L^3 + 5L^2 + \frac{11}{3}L + 1.
\label{eqn:IHN}
\end{equation}
In Fig.\,\ref{fig:2869} we show the atomic configuration for an ideal
Ih structure with the magic number of $N=2869$ atoms ($L=9$).
Atoms are shaded to indicate local fcc, hcp, or other structure.

Different theoretical and numerical models have indicated different limits for the stability of 
such Ih clusters. Using a
continuum approach to take into account the strain energy and the
twin grain boundary energy, Ino predicted that icosahedral clusters should be
stable up to sizes of $40\,000$ atoms \cite{INO}. Similar stability
limits for icosahedral gold clusters were predicted by Marks using a
modified Wulff construction \cite{MARKS_PHIL,MARKS_REVIEW}.  More
recent atomistic calculations\cite{CLEVELAND_ZPD,CLEVELAND_PRL,GARZON}
find that, at $T=0$, icosahedral gold
nanoclusters are the lowest energy structure only in a very small size
range of tens of atoms; other structures, such as an fcc crystal with a truncated octahedral
shape, become energetically favored at larger sizes.
However, in
molecular dynamics simulations of the freezing of gold nanoparticles Chushak
and Bartell \cite{BARTELL} observed icosahedral particles of up to
almost $4000$ atoms. Simulations by Cleveland {\em et al.} \cite{LANDMAN}
found that when fcc truncated-octahedral gold clusters with hundreds of
atoms were heated, they underwent a transformation to the icosahedral structure. 

While the theoretical limit of stability of Ih structures thus remains unclear, 
and their formation may involve kinetic rather than strictly equilibrium effects
\cite{MARKS_REVIEW,INO,CLEVELAND_ZPD,CLEVELAND_PRL,GARZON,MARKS_PHIL}, 
it is natural to suppose that the
formation of the Ih structure is related to the very high stability of 
the $\{111\}$ external surfaces.  
Simulations \cite{CARNEVALI} and experiments \cite{GROSSER} on bulk slab-like 
geometries with exposed $\{111\}$ surfaces have shown that, unlike the 
$\{100\}$ and $\{110\}$ surfaces which melt below the bulk melting 
temperature $T_{\rm m}$, the $\{111\}$ surface is a nonmelting surface 
without roughening, wetting, or melting up to the bulk melting 
temperature $T_{\rm m}$, and can in fact lead to superheating of 
the solid \cite{DITOLLA}. In light of this observation it is interesting
to consider how the high stability of the $\{111\}$ facets effects the
melting and equilibrium shape of such icosahedral clusters. 

In this paper, we will show, by molecular dynamics (MD) simulations, that
liquid gold clusters with about $600$--$3000$ atoms crystallize into 
an Ih structure, with a missing central atom, upon cooling from the liquid. 
We then reheat the clusters back through the melting transition
at temperature $T_{\rm m}$.
Unlike many previous simulations which simulate a continuous heating
process, we simulate heating in quasi equilibrium, running for a long 
simulated time ($43$ ns) at each temperature, before increasing the temperature.
We pay careful attention to the behavior of the cluster surface, and compute for 
the first time the {\em average} cluster shape, as we pass through $T_{\rm m}$.
Using careful measurements of both bond orientational order parameters
and atomic diffusion we find that the $\{111\}$ facets of the cluster surface
stay ordered and do not premelt or roughen below the cluster melting 
temperature $T_{\rm m}$.  Nevertheless, we find that there is a considerable
softening of the cluster surface roughly $\sim 200$ K below $T_{\rm m}$
that can be regarded as due to the ``melting" of the atoms on the vertices and
edges of the cluster.  As temperature increases, there is an increasing mobility of
these atoms leading to intra-layer and inter-layer diffusion, and a shrinkage of the
average area of the $\{111\}$ factes.  The equilibrium shape
progresses from fully faceted, to faceted with rounded edges, to nearly
spherical just below $T_{\rm m}$. Throughout this region, the interior atoms
of the cluster remain essentially perfectly ordered, until $T_{\rm m}$ is
reached.  Our results refine those of earlier simulations of gold 
clusters \cite{LANDMAN,TOSATTI,BARRAT}, which
used measurements of the radial density distribution and the 
observation of surface diffusion as evidence for a general premelting 
of the cluster surface.  A preliminary report of some of our results has been presented
in Ref.\,\onlinecite{WTD}.


\section{METHODS}

\subsection{Simulation model and methods}

On modern computers {\it ab initio} simulation techniques providing an
accurate description of interaction energies can be used to simulate
systems consisting of up to hundreds of atoms \cite{MARX00}. However,
such methods are too computationally demanding to allow long simulation times.
Using less accurate but computationally less
expensive model potentials, such as the embedded atom method
\cite{FOIL86}, the Murrell-Mottram potential \cite{MURR90}, the
Lennard-Jones potential \cite{WALE97}, the Morse potential
\cite{DOYE97}, the many-body Gupta potential \cite{GARZ91}, or the
many-body ``glue'' potential \cite{ERCOLESSI}, one can extend the size
of simulated gold nanoclusters to more than ten thousand atoms. In this
study, we have chosen the many-body ``glue'' potential because it was
found to yield an accurate description of the bulk, defect and surface
properties of gold \cite{ERCOLESSI}. In the ``glue'' model, the
potential energy of a system of $N$ atoms consists of a sum of pair
potentials $\phi$ and a many-body ``glue'' energy $U$,
\begin{equation}
V = \frac{1}{2}\sum_{i}\sum_{j\neq i}\phi\left(r_{ij}\right)
    + \sum_{i}U\left(n_{i}\right).
\label{eqn:GLUE}
\end{equation}
Here the sums run over all particles,
$r_{ij}=\left|{\mathbf r}_{i}-{\mathbf r}_{j}\right|$ is the
interatomic distance between atoms $i$ and $j$, and $\phi(r)$ is the
pair interaction energy. The many-body ``glue'' energy $U(n_{i})$ depends 
on an effective coordination number $n_{i}$ of atom $i$, which is defined by
\begin{equation}
n_{i} = \sum_{j}\rho\left(r_{ij}\right).
\label{eqn:GLUE_N}
\end{equation}
Here $\rho(r)$ is a short-ranged monotonically decreasing function
of the interatomic distance $r$. We use the specific forms for the
pair potential $\phi(r)$, and the glue terms $U(n)$ and $\rho(r)$, as
given by Ercolessi {\it et al.} in Ref.\,\onlinecite{ERCOLESSI}.

We will simulate our gold clusters by treating each atom as a classical particle. 
Newton's equations of motion are integrated using the velocity Verlet algorithm 
 \cite{FRENKEL} with a time step of $\Delta t = 35$ fs. 
Because the many-body ``glue'' potential uses a cutoff, a cell index
method can be used to reduce the computational time \cite{ALLEN}.
In this method, the simulation box is divided into cubic cells with
side lengths larger than the cutoff distance ($3.9$ {\AA} for the gold
glue model). When calculating
energies and forces one considers only interactions between atoms
within the same cell and the neighboring $26$ cells. This approach
reduces the required computation time from order $N^2$ to order $N$.
On a PC equipped with a $1.5$ GHz AMD Athlon CPU and $1$ GB of memory,
$25\,000$ steps can be carried out per CPU-hour for a system of
$2624$ atoms propagating the system for about $100$ ps.

We will refer to each time step $\Delta t$ of the velocity Verlet algorithm
as one basic molecular dynamics (MD) step.  Since each MD step is an 
integration of Newton's equation, it necessarily conserves the 
total energy of the system, and thus simulates a microcanonical ensemble.
To sample instead in
the canonical ensemble, we supplement this basic MD step according to two
different well known methods.
The Andersen thermostat \cite{ANDERSENT} is a method that
mimics a system
coupled to a heat bath with constant temperature. At the end of each
MD step, each particle of the cluster is, with probability $p\sim 0.03\%$,
assigned a new velocity sampled randomly from the Boltzmann
distribution of a given temperature. The Andersen thermostat samples both
configuration and momentum spaces according to the canonical distribution, 
so that the instantaneous total
kinetic energy fluctuates, as is the case for a real physical system. 
However, the Andersen thermostat method does not conserve the total linear and angular
momenta, and so will cause the system's overall position and orientation to drift
over the course of the simulation. This would complicate our
analysis of cluster shape as well as atomic diffusion, since we want to
measure these quantities with respect to coordinates that stay
fixed with respect to the cluster.

This drawback of the Andersen thermostat can be avoided using the  Gaussian isokinetic 
thermostat \cite{EVANS}
which keeps the total kinetic energy $K$ of the system fixed at a value corresponding to a given temperature $T$, $K=(3/2)Nk_BT$. We implement this thermostat by rescaling all velocities by a constant factor after each basic MD step so as to conserve the total kinetic energy. Although the Gaussian isokinetic dynamics does not sample the canonical distribution in momentum space, this method
does correctly sample configuration space and so yields correct equilibrium averages of all structural quantities that depend on positional coordinates only. This ``constant temperature"
method has the advantage of conserving both total linear and angular momenta, thus keeping the cluster at a fixed position and orientation. 

In our simulations, our first goal will be to cool our cluster to low
temperature from a liquid melt to see what solid structure forms. To
do this we use the Andersen thermostat method since we believe it models
more closely the true dynamics of the physical system, and hence will
incorporate effects that may be due to kinetics rather than just pure
thermodynamics. Later, to observe
the equilibrium shape and other properties of the cluster, we will use
the constant temperature MD to heat back through melting. This will keep
the cluster position and orientation fixed, and so simplify our analysis.

\subsection{\label{sec:BOP}Quantifying structure by bond 
            orientational order parameters}

To determine the crystalline structure of our gold nanoclusters, we will
use the method of bond orientational order parameters \cite{STEINHARDT},
which we now review. 
The idea of the bond order parameters is to capture the
symmetry of bond orientations regardless of the bond lengths. 
A bond is defined as the vector joining a pair of neighboring atoms. 
Throughout our paper, 
we will define the neighboring atoms of a given atom $i$
as those atoms which have an interatomic distance less than a cutoff 
radius of $3.8$ {\AA}, equal to the distance to the minimum between
the first and the second peaks of the pair correlation function. 
The local order parameters associated with a bond
${\mathbf r}$ are the set of numbers,
\begin{equation}
Q_{lm}({\mathbf r})\equiv Y_{lm}(\theta({\mathbf r}),\phi({\mathbf r})),
\label{eqn:LOCAL_Q}
\end{equation}
where $\theta({\mathbf r}$) and $\phi({\mathbf r}$) are the polar and
azimuthal angles of the bond with respect to an arbitrary but fixed reference
frame, and $Y_{lm}(\theta({\mathbf r}),\phi({\mathbf r}))$ are the usual
spherical harmonics. 
Since the bond between atoms $i$ and $j$ may be arbitrarily taken as either
${\mathbf r}_{ij}$ or ${\mathbf r}_{ji}$ = $-{\mathbf r}_{ij}$, 
it is useful to consider only the even-$l$ bond
parameters $Q_{lm}$, since only these are invariant to such bond inversions.
Global bond order
parameters can then be calculated by averaging $Q_{lm}({\mathbf r})$
over all bonds in the cluster,
\begin{equation}
\overline{Q}_{lm}\equiv \frac{1}{N_{b}}\sum_{\rm bonds}Q_{lm}({\mathbf r}),
\label{eqn:GLOBAL_Q}
\end{equation}
where $N_{b}$ is the number of bonds. To make the order parameters
invariant with respect to rotations of the reference frame, the
second-order invariants are defined as,
\begin{equation}
Q_{l}\equiv \sqrt{\frac{4\pi}{2l+1}\sum_{m=-l}^{l}
\left| \overline{Q}_{lm}\right|^{2}  },
\label{eqn:QL}
\end{equation}
and the third-order invariants are defined as,
\begin{equation}
W_{l}\equiv
\sum_{\begin{array}{c}
       m_{1},m_{2},m_{3}\\
       m_{1}+m_{2}+m_{3}=0
      \end{array}}
    \left(\begin{array}{ccc}
      l & l & l \\
      m_{1} & m_{2} & m_{3}
     \end{array}\right)
     \overline{Q}_{lm_{1}}\overline{Q}_{lm_{2}}\overline{Q}_{lm_{3}},
\label{eqn:WL}
\end{equation}
where the coefficients $(\cdots)$ are the Wigner $3j$ symbols
\cite{LAND65}. It is standard to define a normalized quantity,
\begin{equation}
\hat{W}_{l}\equiv \frac{W_{l}} {\left( \sum_{m}
                \left| Q_{lm}\right|^{2} \right)^{3/2}}
\label{eqn:HATWL}
\end{equation}
which, for a given $l$, is independent of the magnitudes of the $\{Q_{lm}\}$.

\begin{figure}
\epsfxsize=8.6truecm
\epsfbox{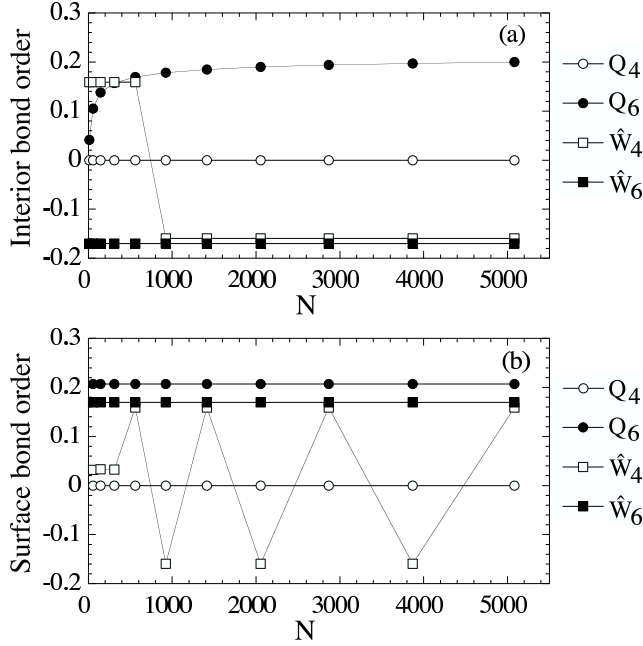}
\caption{The values of (a) the bulk and (b) the surface
bond orientational order parameters
vs. the cluster size $N$, for ideal icosahedral clusters with magic numbers $N$.}
\label{fig:IHBOP}
\end{figure}

The four bond order parameters $Q_4$, $Q_6$, $\hat{W}_4$, $\hat{W}_6$ are 
generally sufficient to identify different crystal structures.  
In Table\,\ref{tab:BOP} we give their values
for ideal periodic fcc, hcp, sc, and bcc crystal structures.  
Since the Ih structure
is not periodic, it may in principle have values for the bond order
parameters that depend on the cluster size.  In Fig.\ref{fig:IHBOP}(a) 
we plot the
values of the four bond orientational order parameters vs. the cluster size
$N$ for several ideal Ih structures.  Although there is a strong size
dependence for small clusters, the bond parameters saturate to well defined values
as $N$ increases, and we list these in Table\,\ref{tab:BOP}.  
Note that although, for large $N$,
most of the atoms in the Ih structure have a local fcc structure, the
values of the bond order parameters are different from those of a pure fcc
crystal; this is due to averaging the order parameters over the differing
orientations of the  $20$ fcc tetrahedra that make up the Ih
structure.

\begin{table}
\caption{\label{tab:BOP}Bond order parameters for face-centered-cubic (fcc),
hexagonal close-packed (hcp), simple cubic (sc), body-centered-cubic (bcc),
liquid, Ih bulk, and Ih surface structures.}
\begin{ruledtabular}
\begin{tabular}{ldddd}
Geometry & Q_4 & Q_6 & \hat{W}_4 & \hat{W}_6
\\ \hline
fcc & 0.190\,94 & 0.574\,52 & -0.159\,32 & -0.013\,16
\\
hcp & 0.097\,22 & 0.484\,76 & 0.134\,10 & -0.012\,44
\\
sc  & 0.763\,76 & 0.353\,55 & 0.159\,32 & 0.013\,16
\\
bcc & 0.082\,02 & 0.500\,83 & 0.159\,32 & 0.013\,16
\\
liquid & 0 & 0 & 0 & 0
\\
Ih bulk & 0 & 0.199\,61 & -0.159\,32 & -0.169\,75
\\
Ih surface & 0 & 0.207\,29 & \pm 0.159\,32 & 0.169\,75
\\ 
\end{tabular}
\end{ruledtabular}
\end{table}

The bond orientational order parameters, averaged over all bonds, will be
used to monitor global structural changes of our cluster.  In particular,
a large liquid cluster will have vanishing values for the bond order
parameters.
Thus the decay of the bond parameters from their 
finite low temperature values to
zero will be a signature of the melting transition.  In this work we will
be particularly concerned with the behavior of atoms on the surface of the
cluster. Surface structures of nanomaterials can often be quite different
from the bulk.  We therefore will consider separately the {\em bulk}
bond order parameters, computed by averaging over only those
bonds connecting atoms that are in the internal ``bulk'' of the
cluster, vs. the {\em surface} bond order parameters, computed by averaging
over only those bonds connecting pairs of atoms that lie on the
surface of the cluster.  In Fig.\,\ref{fig:IHBOP}(b) 
we plot the values of such surface bond
order parameters for an ideal Ih structure vs. cluster size $N$.  Again we
see that they approach well defined constants as $N$ increases, except for
$\hat{W}_4$ which oscillates.  The values of these large $N$ surface bond 
parameters are listed in Table\,\ref{tab:BOP}.  
Note that for the same Ih structure, the
bulk bond parameters have different values than the surface bond parameters,
since they are measuring properties of three dimensional vs. two dimensional
structures, respectively.

\subsection{Geometrical analysis of the surface}

A main goal of this work is to quantify the geometrical behavior of the 
surface of
the gold nanocluster. In this section we describe the algorithms that we
use for this characterization.

\subsubsection{\label{sec:cone}Cone algorithm}

The first step in our analysis is to identify the surface particles of the
clusters accurately according to their geometrical positions. We have
developed a new algorithm, which 
we call the ``cone'' algorithm, to do this.
For a given particle, 
we define an associated {\em cone region} as the region inside a cone of
side length $a$ and angle $\theta$, whose vertex resides on the particle.
A {\em hollow cone} is a cone region with no other
particles inside it. We take a particle to be on the surface if at least 
one associated hollow cone can be found. 

\begin{figure}
\epsfxsize=8.6truecm
\epsfbox{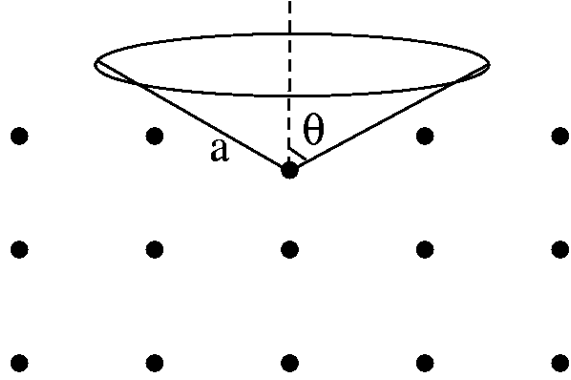}
\caption{Schematic of the cone algorithm.}
\label{fig:CONE}
\end{figure}

\begin{figure}
\epsfxsize=8.6truecm
\epsfbox{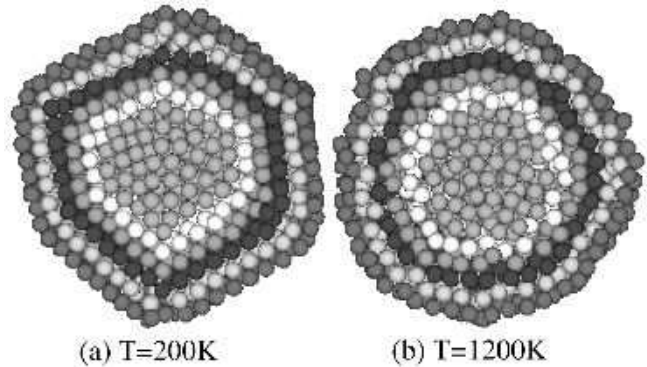}
\caption{Cross section of a gold cluster with $2624$ atoms
at (a) $T=200$ K in an Ih structure and (b) $T=1200$ K in the liquid, 
with the $5$ topmost layers, as determined
by the cone algorithm, marked by different gray scales.}
\label{fig:HALF}
\end{figure}

The cone algorithm is intrinsically consistent with the general definition
of surface particles. It can pick up all of the particles
on a convex surface. For the particles on a concave surface, the precision
of identifying a surface atom relies on the choice of the parameters
$a$ and $\theta$ (see Fig.\,\ref{fig:CONE}). 
By visual examination of our generated
configurations, we found that the two parameters
$a=5.0$ {\AA} and $\theta=\pi/3$ gave good results for our clusters.
For a gold cluster with $5082$ atoms, 
a complete determination of the surface atoms requires less than 2 s of CPU time.
The cone algorithm can also be applied recursively to divide particles 
into surface and sub-surface layers to allow inter-layer analysis. 
Fig.\,\ref{fig:HALF} shows a planar slice cut through an instantaneous 
configuration
of a gold cluster with $2624$ atoms at (a) $T=200$ K in an Ih structure 
and (b) $T=1200$ K in the liquid,
respectively. From this figure we can see that the cone algorithm works
well on both solid and liquid configurations.

\subsubsection{\label{sec:ROUGH}Curvature}

Having identified the surface particles, we next want to quantify the surface 
morphology of the clusters. To do this,  we compute two different measures
of the surface curvature, the {\it bond curvature} and the {\it maximal local
surface curvature}, as defined below.

Our first step is to determine a tangent plane to the cluster surface at
each surface particle. To do this, we consider the collection
of particles determined by the particle of interest and all its
neighboring particles which are also on the surface.  Denote
the coordinates of these particles by ${\bf r}_i\equiv (x_i, y_i, z_i)$,
$i=1\dots N_s$, where $N_s$ is the number of the particles
under consideration.  We define the tangent
plane to pass through the center of mass of these particles, and we
determine its orientation by minimizing the mean square distance
of the particles to the plane. Specifically, 
we solve for this plane as follows.
If $\hat {\mathbf n}\equiv (n_x, n_y, n_z)$ is the unit normal vector of the 
tangent plane
and ${\mathbf r}_{c}\equiv (x_c, y_c, z_c)$ is the coordinate of the center 
of mass, then the distance from the $i$th particle to the tangent plane is,
\begin{equation}
  d_{i} = \hat {\mathbf n}\cdot({\mathbf r}_{i} - {\mathbf r}_{c}).
\label{eqn:DISTANCE}
\end{equation}
We determine the normal vector $\hat {\mathbf n}$ by minimizing 
$\sum_i d_i^2$ subject to the constraint $\hat {\mathbf n}^2 = 1$. Introducing
an undetermined Lagrange multiplier $\lambda$, we solve for,
\begin{equation}
\frac{\delta}{\delta{\mathbf n}}\left[ \sum_i d_i^2
		- \lambda\left(\hat {\mathbf n}^{2} - 1\right)\right] = 0.
\label{eqn:LAMBDA}
\end{equation}
Eq.\,(\ref{eqn:LAMBDA}) leads to the symmetric eigenvalue problem,
\begin{equation}
\left(\begin{array}{ccc}
   XX - \lambda & XY & XZ \\
   XY & YY - \lambda & YZ \\
   XZ & YZ & ZZ - \lambda 
\end{array}\right)
\left(\begin{array}{c} 
   n_x \\
   n_y \\
   n_z
\end{array}\right) = 0,
\label{eqn:TANMX}
\end{equation}
where $XY = \sum_{i}(x_i - x_c)(y_i - y_c)$, and other quantities are
computed similarly.
The smallest of the three possible eigenvalues $\lambda$ determines the 
minimum value of $\sum_i d_i^2$, and its
associated eigenvector $(n_x, n_y, n_z)$ is the normal vector 
$\hat {\mathbf n}$ of the tangent plane. 

\begin{figure}
\epsfxsize=4.2truecm
\epsfbox{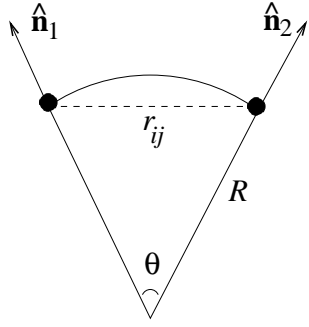}
\caption{Schematic for the calculation of the bond curvature.}
\label{fig:BOND}
\end{figure}

We now define the {\em bond curvature}, $c_b$, of a bond
connecting two neighboring surface particles.
Consider two neighboring surface particles at positions ${\mathbf r}_1$ and ${\mathbf r}_2$
and let $\hat {\mathbf n}_1$ and $\hat {\mathbf n}_2$ be the unit normal vectors
of their corresponding tangent planes.  We can uniquely fit a circle that passes through
the two points, such that $\hat {\mathbf n}_1$ and $\hat {\mathbf n}_2$ are normal 
to the circle.  The bond curvature $c_b$ is then defined in terms of the radius $R$ of this
fitted circle,
\begin{equation}
  c_b \equiv \frac{1}{R} = \frac{2{\sin}(\theta/2)}{r_{ij}},
\label{eqn:BONDCURV}
\end{equation}
where $\theta = {\rm acos}(\hat {\mathbf n}_1\cdot{\hat \mathbf n}_2)$ 
is the angle between the two normal vectors 
and $r_{ij} = \left|{\mathbf r}_1 - {\mathbf r}_2\right|$ 
is the distance between the two particles.
The geometry of this fit is illustrated in Fig.\,\ref{fig:BOND}.

\begin{figure}
\epsfxsize=4.2truecm
\epsfbox{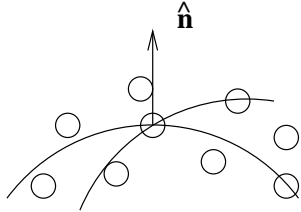}
\caption{Schematic for the calculation of the local surface curvatures.}
\label{fig:PARA}
\end{figure}

Alternatively, we can compute the {\em surface curvature} at each particle
on the surface as follows. Consider a particle on the surface, and all its
neighboring particles that are also on the surface. Since the geometry of these points
roughly defines a surface in $3$D space, we find the best fit of a
paraboloid surface to these points.
The {\em maximal local curvature}, $\kappa_{\rm M}$, and the {\em minimal local curvature},
$\kappa_{\rm m}$, are then given by the two principal
curvatures of the fitted paraboloid. 
The schematic of this method is shown in Fig.\,\ref{fig:PARA}.
To fit the paraboloid at the surface particle ${\mathbf r}_0$, 
we take the normal vector $\hat {\mathbf n}$ to the tangent surface
at ${\mathbf r}_0$, as computed above,
and define this to be the local $z$-axis. We place the origin at
${\mathbf r}_0$ and then choose arbitrary but fixed 
$x$ and $y$ axes in the tangent plane. We
then define the coordinates $(x_i, y_i, z_i)$ of neighboring particle
${\mathbf r}_i$ in this coordinate system. We then choose a
paraboloid function,
\begin{equation}
  f(x,y) = a_{xx}x^{2} + 2a_{xy}xy + a_{yy}y^{2},
\label{eqn:PARAFIT}
\end{equation}
and fit it through the neighboring points ${\mathbf r}_i$ by minimizing
\begin{equation}
  S = \sum_{i}\left(z_{i} - f(x_{i},y_{i})\right)^{2},
\label{eqn:PARAMIN}
\end{equation}
with respect to $a_{xx}$, $a_{xy}$, and $a_{yy}$. This leads to the following
set of linear equations,
\begin{equation}
\left(\begin{array}{ccc}
 \sum_{i}x_{i}^{4}  & \sum_{i}x_{i}^{3}y_{i} & \sum_{i}x_{i}^{2}y_{i}^{2} \\
 \sum_{i}x_{i}^{3}y_{i} & \sum_{i}x_{i}^{2}y_{i}^{2} & \sum_{i}x_{i}y_{i}^{3}\\
 \sum_{i}x_{i}^{2}y_{i}^{2} & \sum_{i}x_{i}y_{i}^{3} & \sum_{i}y_{i}^{4} 
\end{array}\right)
\left(\begin{array}{c} 
   a_{xx} \\
   2a_{xy} \\
   a_{yy}
\end{array}\right) = \left(\begin{array}{c}
   \sum_{i}x_{i}^{2}z_{i} \\
   \sum_{i}x_{i}y_{i}z_{i} \\
   \sum_{i}y_{i}^2z_{i}
\end{array}\right).
\label{eqn:PARAMX}
\end{equation}
Solving Eq.\,(\ref{eqn:PARAMX}) for $a_{xx}$, $a_{xy}$, $a_{yy}$, we 
diagonalize the symmetric
matrix with the elements $a_{\mu \nu}$ to obtain the two 
principal axes and
the corresponding eigenvalues $\lambda_1$ and $\lambda_2$.
The local curvatures are then given by $\kappa_{1,2}=\frac{1}{2}\lambda_{1,2}$.
We will see that the maximal local curvature will be very helpful for
visualizing the vertex, edge and facet atoms of the cluster surface.

\begin{figure}
\epsfxsize=8.6truecm
\epsfbox{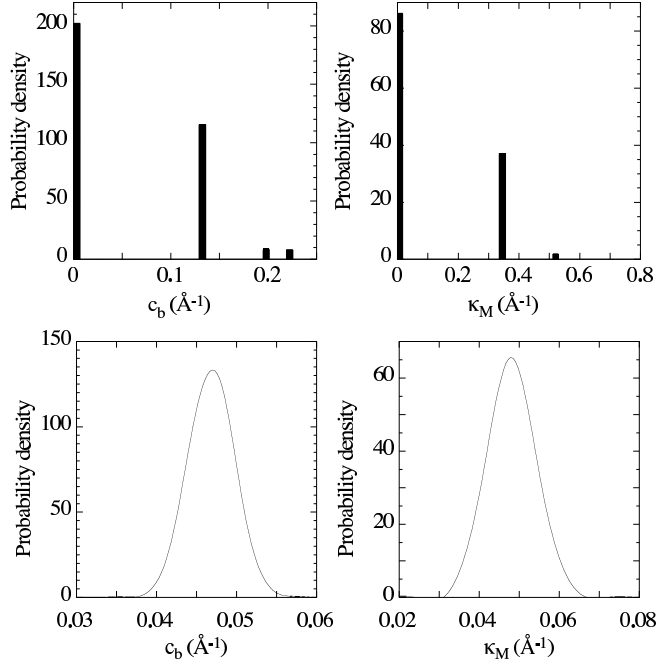}
\caption{Histograms of bond curvature $c_b$ and maximal local curvature 
$\kappa_M$ of (i) top row: an ideal Ih cluster with $N=2869$ atoms,
and (ii) bottom row: the average shape of a liquid cluster with $N=2624$ atoms.}
\label{fig:IDEAL}
\end{figure}

To test these two methods, we consider the ideal
Ih gold cluster with a magic number of atoms $N=2869$, shown in
Fig.\,\ref{fig:2869}. In Fig.\,\ref{fig:IDEAL} (top row) we show the
resulting histograms of bond curvature and maximal local curvature for this cluster.
Both histograms consist solely of  $\delta$ functions, corresponding to the facets,
the edges and the vertices of the Ih structure. For the bond curvature
histogram there are {\em two} separate peaks for the vertices,
corresponding to the bond curvatures between the vertex atoms and the edge
atoms, and the vertex atoms and the facet atoms. We also test
our method for the average shape of 
a liquid gold cluster with $2624$ atoms at $T=1200$ K (shown in
Fig.\,\ref{fig:DIST}(c)). We show our results in Fig.\,\ref{fig:IDEAL} (bottom
row). Both histograms have just one peak of finite width,
centered at $1/R = 0.047$ {\AA}$^{-1}$, with
$R = 21.5$ {\AA} the radius of the spherical liquid drop.
Note that while the average shape of the liquid drop is close to a perfect sphere,
the histogram of its curvatures seems relatively broad.  We have found that this is
due to the small discrete number of neighboring surface particles that is used
to define the fitted paraboloid.  Even small deviations of these particles from a
constant radius can lead to noticeable variations in the fitted curvatures.  We find that we
can reduce the width of the curvature histogram for the liquid cluster 
by increasing the cutoff length used to define neighboring particles, and so have
more particles included in the fits to the local paraboloids.  However, while this
improves the calculation for a spherical cluster, it makes it worse for high curvature
regions near edges and vertices in an Ih cluster, where curvature varies rapidly
as one moves across the surface. With this understanding of 
our method's limitations, we therefore leave our cutoff as 
given above.

\subsubsection{\label{sec:AVERAGE}Average shape}

At low temperatures, the atoms in the gold cluster remain at well defined 
equilibrium positions and only thermally oscillate around the vicinity of these
equilibrium positions. The shape of the cluster is thus easily determined from
an instantaneous configuration.
At high temperatures, however, atoms become more mobile
and the macroscopic shape of the cluster fluctuates dramatically from
configuration to configuration. In this case it becomes necessary to average 
over many fluctuating configurations to define the average cluster shape.
Since our constant temperature MD conserves total linear and angular
momenta and both are set to zero, the configurational shape changes we 
average over represent
fluctuations of the surface atoms rather than trivial shifts or rotations
of the cluster as a whole.

At high temperatures, simply averaging the position of each atom throughout 
the course of the simulation does not give the 
average shape because the atoms are in general 
no longer bound to specific sites but may diffuse many interatomic spacings
through the cluster. 
We therefore use the following approach to define the average cluster shape.
We divide the surface of the cluster into equal solid angles, 
and then average the instantaneous surface atom positions in each solid angle. 
This average position in each solid angle then defines the profile of the
cluster's average shape. This average shape does not
contain information about the 
individual surface atom positions, since generally a given
solid angle may contain the instantaneous positions of different surface atoms
at different times. To define our solid angle division,
we use the best covering spherical codes 
with icosahedral symmetry \cite{HARDIN} to
divide the $4\pi$ total solid angle of the sphere centered at the center 
of mass into cone cells with almost equal solid angles. 
Choosing different numbers of solid angles results in different resolutions;
we always 
choose a number of solid angles that matches as close as possible to 
the number of surface atoms in the cluster.

\begin{figure}
\epsfxsize=8.6truecm
\epsfbox{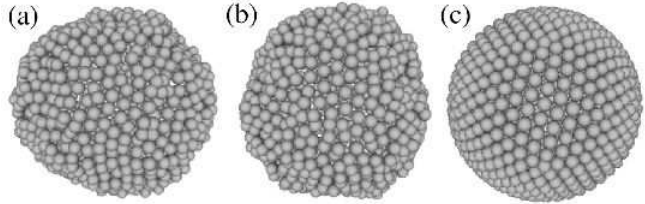}
\caption{A liquid gold cluster with $2624$ atoms
at $T=1200$ K. (a) and (b) are two instantaneous configurations. 
(c) is the shape averaged over $1000$ such instantaneous configurations.}
\label{fig:DIST}
\end{figure}

We illustrate this method for a liquid gold cluster with $2624$ atoms
at $T=1200$ K. In Figs.\,\ref{fig:DIST}(a) and (b) we show two arbitrary 
instantaneous configurations of the cluster. 
Theoretically, a liquid cluster should have a perfect sphere as the 
equilibrium shape. However,
we see that the instantaneous configurations can have noticeably
large fluctuations about the average shape. Applying our shape averaging
procedure above on $1000$ configurations sampled at equal time intervals 
from $43$ ns of simulated time, we recover that 
the average shape, shown in Fig.\,\ref{fig:DIST}(c), is
a perfect sphere as expected. Note that in Figs.\,\ref{fig:DIST}(a) and (b),
the small spheres represent instantaneous atomic positions. In contrast, 
in Fig.\ref{fig:DIST}(c), they represent not specific atoms, but rather
the average surface position within the given solid angle.

\subsection{\label{sec:DIFF}Atom diffusion analysis}

With enough kinetic energy, atoms can hop around their crystal sites and even
travel across the whole cluster. The mean squared displacement (MSD)\cite{FRENKEL}, 
$\Delta r^2(t)$,  is a 
convenient way to measure the average movement of a group of 
atoms. It is defined as
\begin{equation}
  \Delta r^2(t) = \frac{1}{MN_s}\sum_{j=1}^{M}\sum_{i=1}^{N_s}
                   \left[ {\mathbf r}_{i}(t_{j}+ t) - 
			  {\mathbf r}_{i}(t_{j}) \right]^{2},
\label{eqn:MSD}
\end{equation}
where ${\mathbf r}_i$, $i=1\ldots N_s$ are the positions of the $N_s$ atoms
under consideration, and $t$ is the time interval over which the motion takes place.
We average over $M$ non-overlapping time intervals, 
with $t_j\equiv t_{j-1}+t$.
For an infinite three dimensional bulk system, we expect that 
$\Delta r^2=6Dt$ as
$t\rightarrow\infty$, where $D$ is the diffusion coefficient. 
In a finite cluster, however, the MSD will eventually
saturate on a length scale comparable to the cluster size. We therefore
determine the diffusion coefficient $D$ by fitting 
$\Delta r^2(t)$ to the
early time linear part before saturation takes place.

We will also find that a convenient way to visualize individual atomic displacements
is through an {\em ellipsoid of displacement}. We compute this ellipsoid as
follows. For a given atom traced through $K$ 
successive configurations for a simulation time $t$, 
the mean squared displacement correlations are
given by the $3\times3$ matrix $\mathbf{C}$ with elements,
\begin{equation}
  C_{\mu\nu} \equiv \frac{1}{K}\sum_{i=1}^{K}
	(r_{i\mu} - \langle r_{\mu}\rangle)(r_{i\nu} - \langle r_{\nu}\rangle),
\label{egn:elmat}
\end{equation}
where $\mu,\nu=x, y, z$, $r_{i\mu}$ is the $\mu$-coordinate of the atom in
configuration $i$,
and $\langle r_{\mu}\rangle$ is the average of the coordinate 
over all $K$ configurations. The probability for the atom 
to be at position ${\mathbf r}$ is then approximated as 
$P({\mathbf r})\sim \exp (-\frac{1}{2}[({\mathbf r}-\langle{\mathbf r}\rangle)
\cdot {\mathbf C}^{-1} \cdot ({\mathbf r}-\langle{\mathbf r}\rangle)])$, 
and so the surface of our ellipsoid of displacement is given by the equation,
\begin{equation}
({\mathbf r}-\langle{\mathbf r}\rangle) \cdot {\mathbf C}^{-1} \cdot
({\mathbf r}-\langle{\mathbf r}\rangle) = 1.
\label{eqn:elsurf}
\end{equation}
The eigenvectors of $C_{\mu\nu}$ and the square root of their
corresponding eigenvalues then define the axes and principal radii of the
ellipsoid, which we center on the average atom position $\langle {\mathbf r}\rangle$. 
This ellipsoid provides
a convenient visualization of the directional distribution of root
mean squared displacements over the time $t$.


\section{RESULTS}

In this section we report on our results.
Gold clusters with more than $5000$ atoms require too much computational 
time to allow for the long simulation times we want in order to explore the
equilibrium behavior. Clusters with fewer than a few hundred atoms,
however, have large
finite size effects due to the larger surface-to-volume ratio. 
Such smaller clusters can undergo
transitions between several different 
crystal structures even at low temperatures\cite{LANDMAN,BARRAT,WANG_DELLAGO},
and they have less sharply defined melting transitions.
In this work we have therefore simulated several clusters in the range of
$600$ to $5000$ atoms.
In our results below, we will concentrate on the moderate size 
of $N=2624$ atoms, for which we have done our most complete and careful
analysis. We will also give less detailed results for two smaller
sizes, $N=603$ and $N=1409$, in order to illustrate general trends.
Note that these values of $N$ are {\em not} among the magic numbers
(see Eq.\,(\ref{eqn:IHN})) needed to construct a perfect Mackay icosahedron. 
Nevertheless we will show that these clusters still form Ih structures
upon cooling. We have also studied several clusters $with$ a magic
number of atoms, by explicitly constructing the Mackay icosahedron at low
temperature, and heating through melting.
We will give results for sizes\cite{missingatom} $N=922$ and $N=5082$ in order to compare
with the other more generic values of $N$.

\subsection{\label{sec:missingatom}Mackay icosahedra with a missing central atom}

Our initial goal is to cool a liquid cluster through the melting transition
to determine the ordered structure into which it solidifies.
We therefore started with a liquid gold cluster with $N=2624$ 
atoms which we roughly equilibrated
at $1500$ K, before cooling to $1200$ K where we equilibrated longer. 
We then cooled the cluster down to $200$ K, decreasing the temperature in
intervals of $100$ K. At each temperature the
system is equilibrated for $5\times10^6$ steps ($21.5$ ns) using
the Andersen thermostat method.
With this cooling method we find that our cluster solidifies into 
an Ih structure\cite{MACKAY}. 

\begin{figure}
\epsfxsize=8.6truecm
\epsfbox{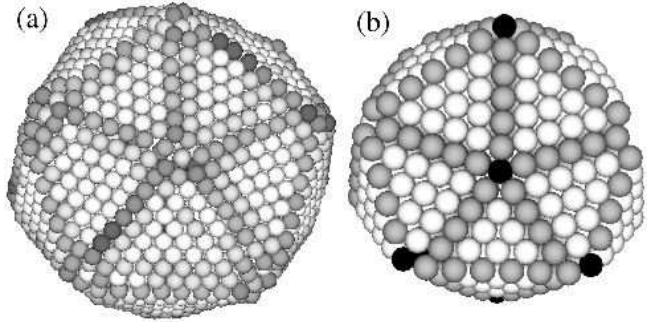}
\caption{Ih structure of an $N=2624$ atom gold cluster at $T=200$ K.
(a) Surface of an instantaneous configuration with atoms
shaded according to the maximal local 
curvature; the larger the curvature, the darker the gray scale. 
(b) The same configuration
with the three outer most layers peeled away. Atoms are shaded according to
their local crystal structure; white is fcc, gray is hcp, and black
is ``other''.}
\label{fig:IH}
\end{figure}

In Fig.\,\ref{fig:IH} we show an instantaneous configuration of this
$N=2624$ gold cluster at our lowest temperature, $T=200$ K. 
To clarify the geometry of the cluster, we have calculated the local 
curvatures for each surface atom according to the method of 
Section\,\ref{sec:ROUGH}, and in Fig.\,\ref{fig:IH}(a) we shade each atom 
according to the 
maximal local curvature; the greater the curvature, the darker the gray scale.
Comparison with Fig.\,\ref{fig:2869} strongly suggests that our cluster has
an Ih structure. Large curvature regions correspond to edges and vertices,
while low curvature regions are the flat $\{111\}$ facets of the fcc
tetrahedra. Note that some vertices have low curvature; this
is because these vertices have their top most atom missing, and so
form a small locally flat region.

To further illustrate the Ih nature of our cluster, we have computed the 
local bond order parameters for each atom, averaging over all bonds that
connect the given atom to its neighbors.
Using the values in Table\,\ref{tab:BOP}, we then identify each atom
with its local crystal structure. We regard atoms
with $Q_4 > 0.15$ and $\hat W_4 \leq 0$ as having a local fcc structure,
and atoms with $Q_4 \leq 0.15$ and $\hat W_4 > 0$ as having a local hcp
structure; all other atoms are simply labeled as ``other''. 
Because the surface layer and the two sub layers 
closest to the surface exhibit surface reconstruction and have
frozen in surface
fluctuations, we have peeled them away by use of the cone algorithm
of Section\,\ref{sec:cone}.
The resulting interior of the cluster is shown in Fig.\,\ref{fig:IH}(b),
where fcc atoms are shaded
white, hcp atoms gray, and ``other'' atoms black.
The Ih structure of the cluster is readily apparent. One clearly sees
the flat $\{111\}$ facets of the fcc tetrahedra, the edges of the facets
corresponding to the hcp twin grain boundaries, and the vertices with
$5$-fold symmetry. 

\begin{figure}
\epsfxsize=8.6truecm
\epsfbox{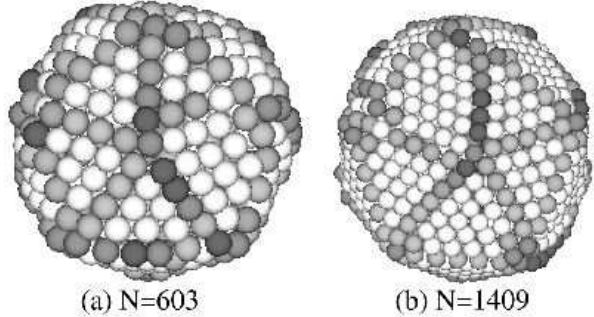}
\caption{Ih structure of gold clusters with
(a) $N=603$ and (b) $N=1409$ atoms at $T=200$ K.
The atoms on the surfaces of these instantaneous configurations are shaded 
according to the maximal local curvature; the larger the curvature,
the darker the gray scale.}
\label{fig:OTHERIH}
\end{figure}

We have also applied the same cooling procedure on smaller gold clusters
with $N=603$ and $N=1409$ atoms.
In Fig.\,\ref{fig:OTHERIH} we show the instantaneous configurations of
$N=603$ and $N=1409$ at $T=200$ K, with surface atoms
shaded by their maximal local curvature (as was done in Fig.\,\ref{fig:IH}(a)
for $N=2624$). We again clearly see the Ih structure, however for the smaller
cluster, the edges and facets appear slightly rounded.

It is interesting to note 
in Figs.\,\ref{fig:IH} and \ref{fig:OTHERIH} that
the fcc tetrahedra
of our clusters are not all of equal size. For a non-magic number $N$ of
atoms, such as is the case here, this is to be expected.
However, we have also cooled clusters with magic numbers\cite{missingatom} $N=560$ 
and $N=1414$ from liquid to $200$ K using the exact same cooling
procedure. These clusters also formed asymmetric
Ih structures with $20$ facets of slightly unequal sizes. This suggests that our
cooling procedure, while slow enough to balance surface vs. bulk free
energy and find the Ih structures, is not slow enough to achieve the
perfect global equilibration which one expects would result in perfectly
symmetric structures for magic numbers $N$.

An interesting feature of our clusters that can not be seen in 
Figs.\,\ref{fig:IH} and \ref{fig:OTHERIH} 
is that all of our clusters formed with a missing central atom.
The energetics of such vacancies at the center of Ih
clusters were first considered by Boyer and Broughton \cite{BOYER}
for Lennard-Jones clusters and later by Mottet {\em et al.}
\cite{LEGRAND} for Cu, Ag, and Au particles. Above a certain
material dependent critical size the central vacancy lowers the
energy of the cluster by partially releasing the strain caused by the
mismatch of the tetrahedral units. Mottet {\em et al.} concluded that
for gold particles the introduction of the central point defect does
not lower the energy enough to make the icosahedron competitive with
crystallographic octahedra and Wulf polyhedra. Their conclusion,
however, was based solely on energy calculations which neglect the entropic
contributions to the free energy at finite temperature. 
Our finite temperature simulations therefore suggest that such
a constitutional vacancy can in fact
stabilize icosahedral clusters of thousands of atoms,
making them the observed structure upon cooling.

\subsection{Melting and the bond orientational order parameters}

Having determined that our clusters cooled from the liquid
have the Ih structure,
we then heated up the clusters using constant temperature MD instead of
the Andersen thermostat, so that the total linear and angular momenta are 
conserved and vanish; this ensures that our clusters neither translate nor
rotate during the course of our simulations. We heat in
temperature intervals of $100$ K when far from $T_{\rm m}$, but use
smaller intervals when approaching $T_{\rm m}$. 
At each temperature the clusters have been equilibrated for $10^6$ MD 
steps ($4.3$ ns), followed by $10^7$ steps ($43$ ns) to collect data. 
Our simulation times are more than an order of magnitude longer than
the $\sim 1$ ns typically simulated in earlier works\cite{LANDMAN,BARTELL}.

\begin{figure}
\epsfxsize=8.6truecm
\epsfbox{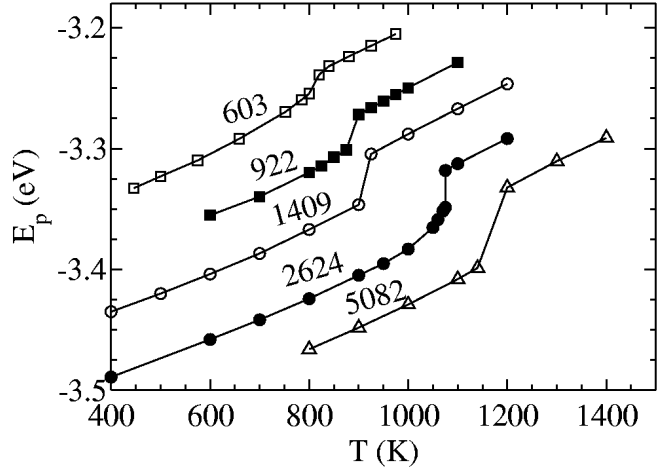}
\caption{Caloric curve of Ih gold clusters with
$N=603$, $1409$, and $2624$ atoms, as well as with magic numbers\protect\cite{missingatom} 
of $N=922$ and $5082$ atoms.}
\label{fig:EP}
\end{figure}

In Fig.\,\ref{fig:EP} we show the caloric curve (average
potential energy per atom vs. temperature) for several of 
our cluster sizes, upon heating. The kink in each curve locates the cluster melting
transition. Several expected trends \cite{TOSATTI} are clearly seen: 
(i) the melting temperature increases as the cluster size increases, 
and (ii) the average potential energy per atom 
increases as the cluster size decreases, due to the larger surface-to-volume ratio.
No qualitative difference is seen between the magic number sizes, $N=922$ and
$5082$, and the others. Note that the glue model gives a melting temperature
of $1357$ K for bulk gold, well above that of our biggest cluster 
\cite{ERCOLESSI}.
The experimentally measured melting temperature of bulk gold 
is $1337$ K \cite{OCKO}.

We have done our most careful heating for the $N=2624$ atom cluster, taking
fine temperature increments near $T_{\rm m}$.  
Heating at the above rate of $43$ ns per temperature, 
we find that the cluster has a first order melting transition at $T=1075$ K. 
However, when we simulated the cluster at the slightly lower temperature
of $T=1070$ K for more than $240$ ns, we found that it also ultimately melted. 
Thus the estimates of $T_{\rm m}$ from Fig.\,\ref{fig:EP} are most likely slightly
higher than the true equilibrium values.  This superheating that we find is 
perhaps related to the extraordinary stability 
of the gold $\{111\}$ surface, as was also observed in a slab-like 
geometry \cite{DITOLLA}. 

Next we wish to explore the melting transition from the perspective
of the bond orientational order parameters, defined in Section\,\ref{sec:BOP}.
We are interested specifically to consider the behavior of the surface of the 
cluster as distinct from the behavior of the interior.
We therefore use the cone algorithm recursively to group the
atoms of the cluster into successive layers. 
The outer most layer of atoms is identified as the surface layer; the atoms
immediately below the surface layer are called the first sub layer, then the
second sub layer, and so on. For the cluster of $N=2624$ atoms 
there are a total of $9$ such layers.
We label the atoms lying below the fourth sub layer
as ``interior'' or ``bulk'' atoms. For $N=2624$, we show in 
Table\,\ref{tab:NUMBER} the number of atoms in each layer for various
temperatures up through melting. What is immediately apparent is that 
as the temperature varies within the solid phase, $T<T_{\rm m}\simeq 1075$ K, 
the number of atoms in a given
layer remains essentially constant, within about $\sim 5$, for all layers below
the second sub layer.  The surface and top two sub layers, however, display
a more noticeable variation, suggesting changes on the surface of the cluster
well below melting. 

\begin{table}
\caption{\label{tab:NUMBER}Average numbers of atoms in the surface layer, 
the sub layers and 
the bulk of an $N=2624$ atom gold cluster at different temperatures.}
\begin{ruledtabular}
\begin{tabular}{ldddddd}
$T$ & \multicolumn{1}{c}{Surface} & \multicolumn{1}{c}{Sub layer 1}
    & \multicolumn{1}{c}{Sub layer 2} & \multicolumn{1}{c}{Sub layer 3} 
    & \multicolumn{1}{c}{Sub layer 4} & \multicolumn{1}{c}{Bulk}
\\ \hline
$400$ K  & 858.5\pm 0.6 & 602.8\pm 0.8 & 428.3\pm 1.1 
         & 307.4\pm 1.1 & 207.2\pm 0.8 & 219.6\pm 1.1 
\\
$600$ K  & 859.8\pm 1.2 & 602.2\pm 1.4 & 427.9\pm 1.2 
         & 307.3\pm 1.1 & 207.2\pm 0.9 & 219.7\pm 1.1 
\\
$900$ K  & 867.7\pm 2.4 & 594.9\pm 2.6 & 427.5\pm 1.4 
         & 307.0\pm 1.2 & 207.0\pm 1.0 & 219.9\pm 1.1 
\\
$1060$ K & 869.9\pm 3.6 & 582.4\pm 4.0 & 436.2\pm 3.2 
         & 311.3\pm 2.6 & 208.6\pm 2.2 & 215.7\pm 3.2 
\\
$1100$ K & 874.7\pm 3.9 & 572.4\pm 4.2 & 436.2\pm 4.2 
         & 308.7\pm 4.0 & 209.9\pm 3.7 & 222.1\pm 5.1 
\\
\end{tabular}
\end{ruledtabular}
\end{table}

\begin{figure}
\epsfxsize=8.6truecm
\epsfbox{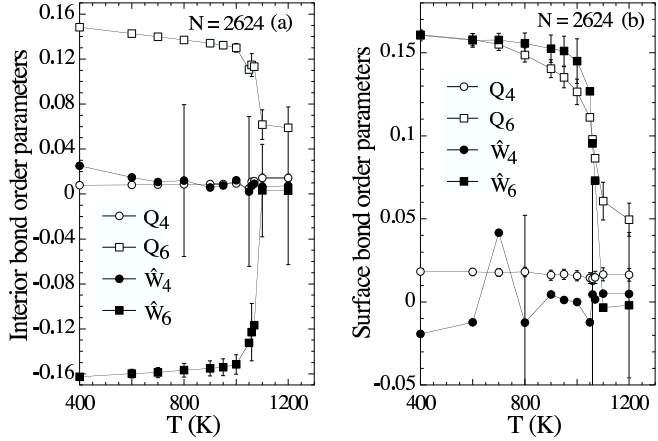}
\caption{Bond orientational order parameters of the $N=2624$ atom cluster
for (a) the interior atoms, and (b) the surface atoms. Sample error bars, representing
configuration to configuration fluctuations, are shown.}
\label{fig:BOP}
\end{figure}

Having made this division into layers,
we then compute the four bond orientational order parameters 
$Q_4$, $Q_6$, $\hat{W}_4$, and $\hat{W}_6$, 
defined in Section\,\ref{sec:BOP}, separately for each layer and for
the bulk. In Fig.\,\ref{fig:BOP} we show our results for the $N=2624$ atom
cluster; Fig.\,\ref{fig:BOP}(a) is for the interior atoms, while 
Fig.\,\ref{fig:BOP}(b) is for the surface atoms. 
Comparing to the values listed in Table\,\ref{tab:BOP}, or equivalently
as shown in Fig.\,\ref{fig:IHBOP}, we see that the values we now find at low
temperature are quite consistent with the bulk and surface values
appropriate for an Ih structure. The only exception to this is the case
of $\hat W_4$ which we find to be approximately zero, rather than the
negative or positive number shown in Table\,\ref{tab:BOP}.
However we have found that, unlike the other bond order parameters, the value of
$\hat W_4$ is extremely sensitive to the symmetry of the perfect Ih structure.
For deviations from this perfect structure, as is the case for our simulated 
cluster, $\hat W_4$ can vary dramatically.  This is evidenced by the very 
large sample to sample fluctuations we found for $\hat W_4$, as indicated by 
the very large error bars shown in Fig.\,\ref{fig:BOP} for $\hat W_4$ as 
compared to the other quantities.
We thus conclude that the bond orientational order parameters are very
consistent with our cluster being a Mackay icosahedron.

In Fig.\,\ref{fig:BOP}(a) for the interior atoms, we see that bulk bond
orientational order parameters remain roughly constant until just above
$1000$ K, before taking a sharp drop towards zero at the same melting 
temperature, $T_{\rm m} \simeq 1075$ K, as found from the caloric curve 
of Fig.\,\ref{fig:EP}.
Thus the bond orientational order parameters give a good signature of the
melting transition. The sharp decrease of the bond parameters indicates
that the interior atoms remain with a highly ordered Ih structure until just
before melting.
Note that the values in the liquid above $T_{\rm m}$ are not identically zero,
but have small finite values due to the finite size of the liquid cluster; this
effect is biggest for $Q_6$.  

In Fig.\,\ref{fig:BOP}(b) for the surface atoms, we again see that
the bond orientational order parameters remain with their Ih values
at low temperatures, and then vanish towards zero at the {\it same} $T_{\rm m}$ as
for the bulk atoms.  Thus we reach one of our most important conclusions:  
the presence of finite surface bond orientational order up until the bulk 
melting transition indicates that the surface $\{111\}$ facets of the Ih structure 
do {\it not} premelt, but rather
{\it the surface facets melt at the same temperature as the bulk}.
The absence of any sharp features in the surface bond orientational order
parameters below $T_{\rm m}$ suggests that there are no other types of surface 
phase transitions below $T_{\rm m}$.
There is, however, one noticeable difference in the behavior of the surface
bond orientational order parameters as compared to the bulk.  We see that 
the surface parameter $Q_6$ starts a noticeable decrease from its low 
temperature
value at $T\sim 800$ K, considerably below the melting $T_{\rm m}$.  We interpret
this as a softening of the surface, and we will 
present the reason for this behavior in the following section.

\begin{figure}
\epsfxsize=8.6truecm
\epsfbox{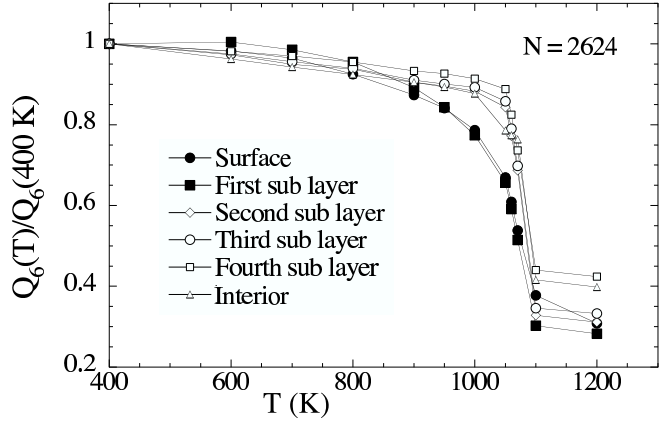}
\caption{Normalized bond orientational order parameter $Q_6(T)/Q_6(400\,{\rm K})$
for the surface, interior, and various sub layers of the $N=2624$ atom cluster.}
\label{fig:q6}
\end{figure}

We have also measured the bond orientational order parameters for the first 
through fourth sub layers of the cluster.  Since  $Q_6$ is the  bond parameter
that most clearly shows the surface softening, in Fig.\,\ref{fig:q6} we plot
the value of $Q_6$ vs. temperature, for surface, interior, and each of the
four sub layers.  Since each layer has a slightly different value of $Q_6$ at
low temperature, we plot the normalized values $Q_6(T)/Q_6(400\, {\rm K})$
so as to better compare their relative behaviors.  We see that both the surface
and the first sub layer show almost identical softening as $T_{\rm m}$ is
approached.  However all the deeper sub layers show almost identical
behavior as the interior atoms, with almost no softening until $T_{\rm m}$.
Thus the softening phenomenon is seen to be largely confined to the top
two layers of the cluster and does not propagate more deeply as 
$T_{\rm m}$ is approached; below the top two layers, the cluster remains
almost as ordered as at low temperatures, until just before melting.

We have also tested the sensitivity of our
definition of the ``interior" atoms of the cluster, by redefining it to 
be {\em all} the atoms below the surface layer.  However, as might be
expected from Fig.\,\ref{fig:q6}, computing 
the bulk bond orientational order parameters defined this way gives no 
qualitative change from the behavior seen in Fig.\,\ref{fig:BOP}(a).

\begin{figure}
\epsfxsize=8.6truecm
\epsfbox{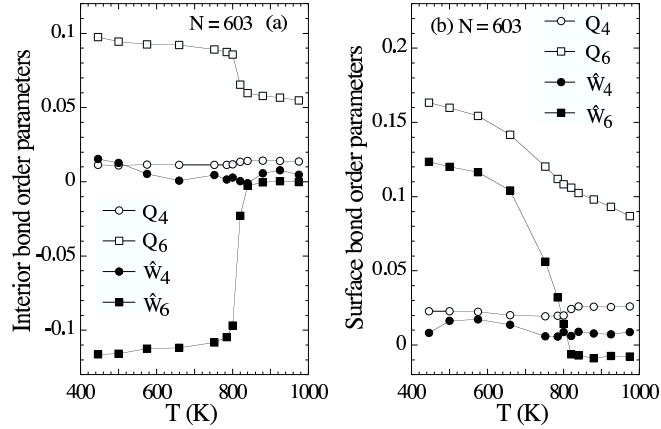}
\caption{Bond orientational order parameters of the $N=603$ atom cluster
for (a) the interior atoms, and (b) the surface atoms.}
\label{fig:603BOP}
\end{figure}

\begin{figure}
\epsfxsize=8.6truecm
\epsfbox{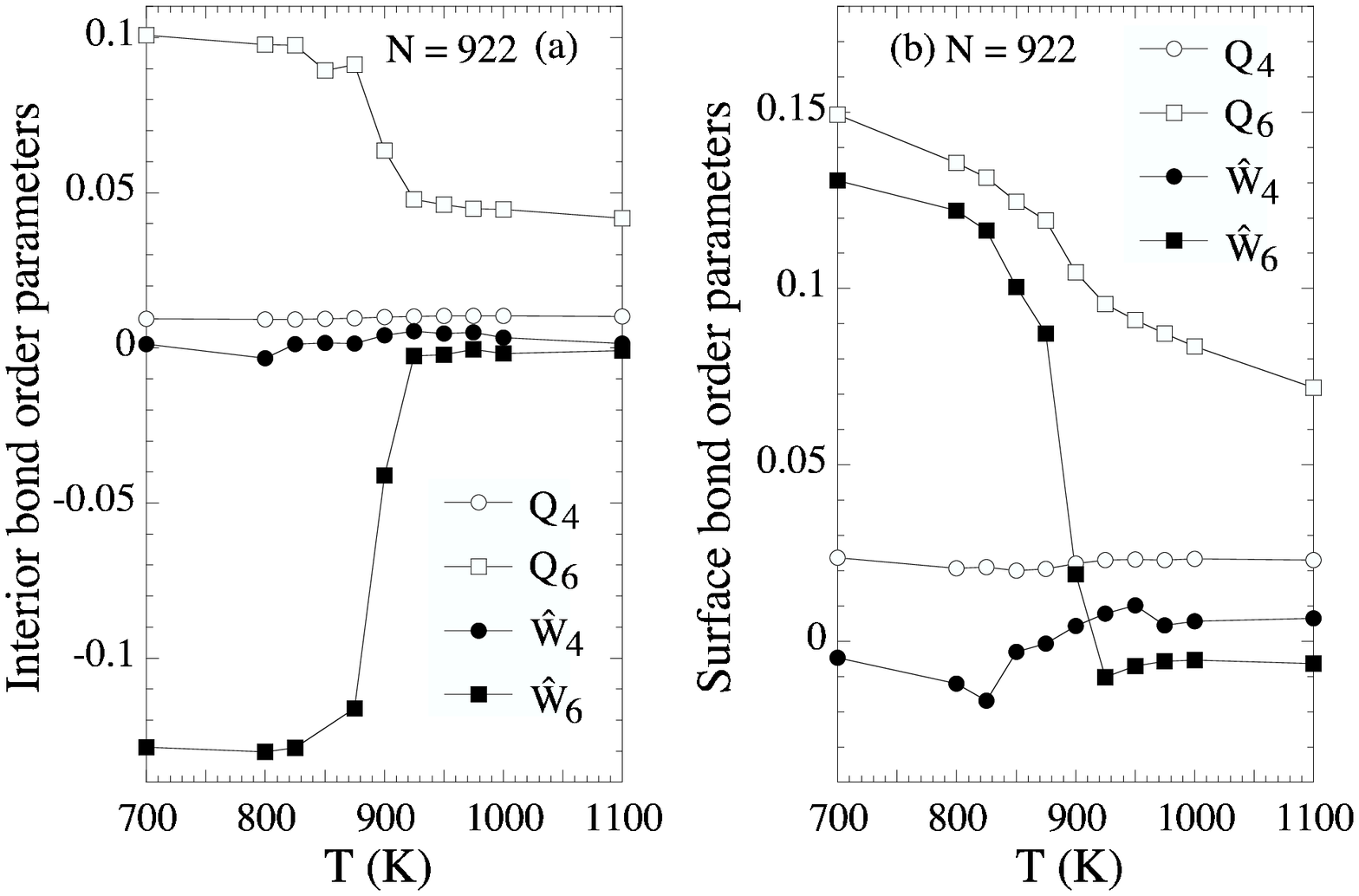}
\caption{Bond orientational order parameters of the magic number $N=922$ atom 
cluster for (a) the interior atoms, and (b) the surface atoms.}
\label{fig:922BOP}
\end{figure}

\begin{figure}
\epsfxsize=8.6truecm
\epsfbox{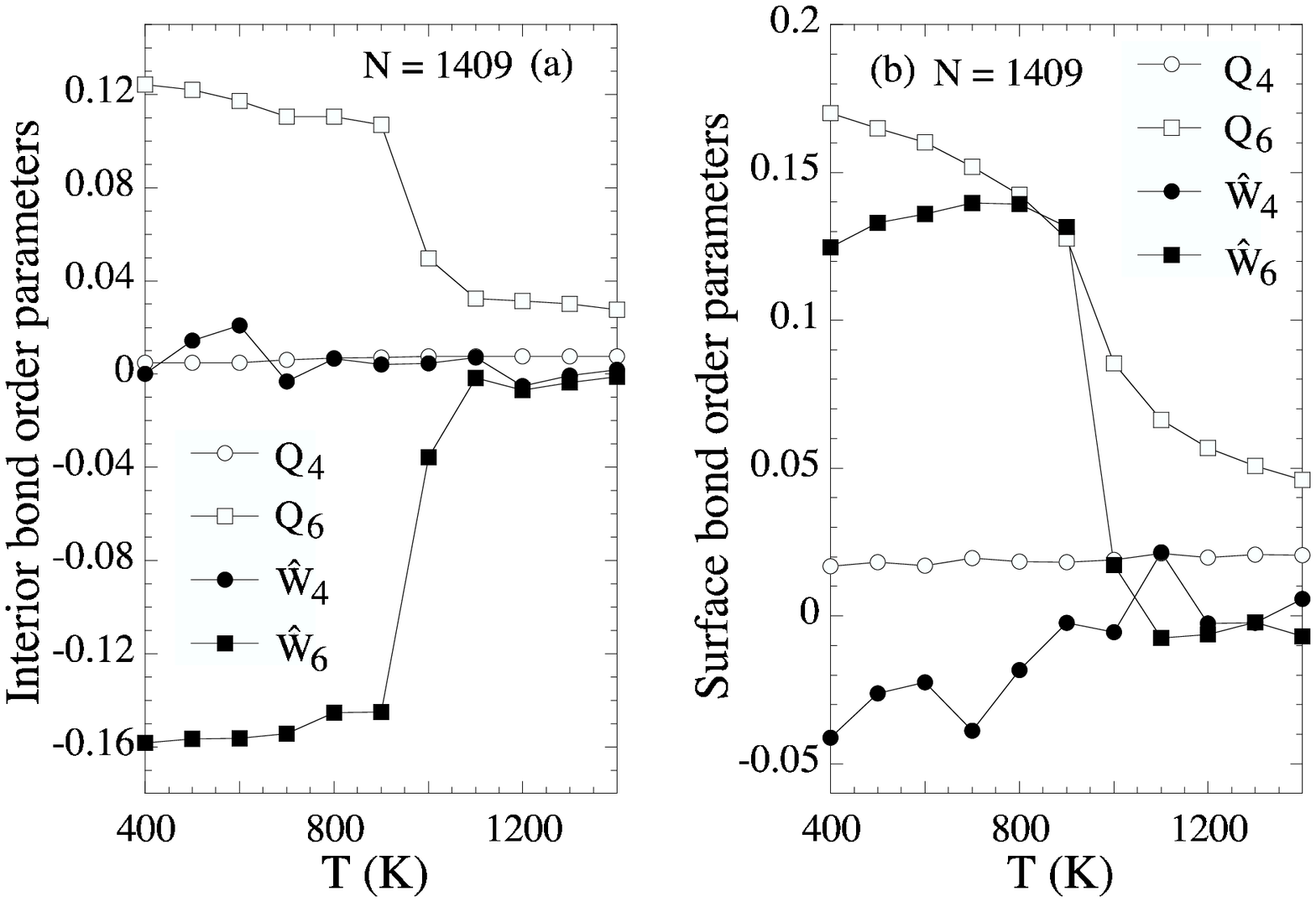}
\caption{Bond orientational order parameters of the $N=1409$ atom cluster
for (a) the interior atoms, and (b) the surface atoms.}
\label{fig:1409BOP}
\end{figure}

\begin{figure}
\epsfxsize=8.6truecm
\epsfbox{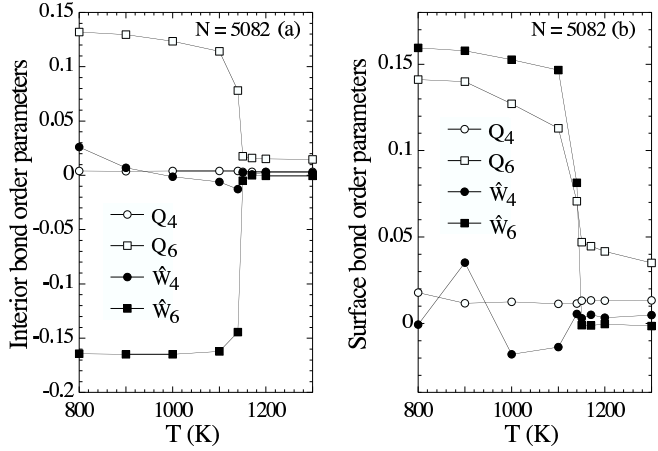}
\caption{Bond orientational order parameters of the magic number $N=5082$ 
atom cluster for (a) the interior atoms, and (b) the surface atoms.}
\label{fig:5082BOP}
\end{figure}

In Figs.\,\ref{fig:603BOP} to \ref{fig:5082BOP} we show similar plots of 
interior and
surface bond orientational order parameters for our other cluster sizes,
$N=603$, $922$, $1409$ and $5082$.  We see the same qualitative
behaviors as in Fig.\,\ref{fig:BOP}, with surface and bulk melting at the
same temperature.  This melting temperature, which increases with cluster
size, agrees with the values found from the caloric curves of Fig.\,\ref{fig:EP}.  
Surface {\em softening} tracks the melting temperature and
starts to be noticeable about $200$ K below $T_{\rm m}$.
The surface softening is somewhat enhanced for
the smaller cluster sizes.  There appears to be no qualitative differences for 
our magic number\cite{missingatom} clusters, $N=922$ and $5082$, 
as compared to the other sizes.

\subsection{Average shape and surface curvature}

\begin{widetext}
\begin{figure}
\epsfxsize=17.2truecm
\epsfbox{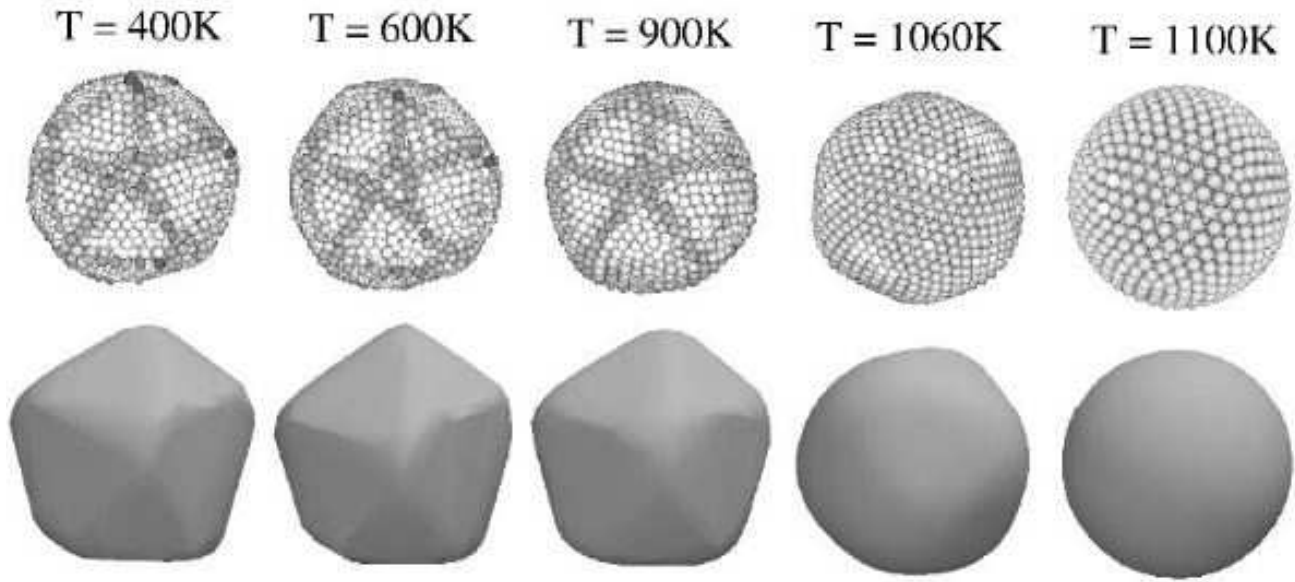}
\caption{Average shapes of an $N=2624$ atom cluster
at $400$, $600$, $900$, $1060$ and $1100$ K.
The top row shows each of the discretized solid angles of the surface shaded
according to the value of the maximal local curvature; the darker
the gray scale, the larger the curvature.  The viewpoint of these
pictures is set to infinity, to show a full hemisphere of solid angle.
The bottom row is the corresponding smooth contour plot, with a finite
viewpoint so as to highlight the straight edges and $5$-fold symmetry about the vertices.}
\label{fig:AVERAGE}
\end{figure}
\end{widetext}

To understand the physical manifestations of the surface
softening that is indicated by the surface bond orientational
order parameters, we now look at the average shapes
of our cluster, computed according to the method of
Section\,\ref{sec:AVERAGE}.  We focus first on our cluster of $N=2624$ atoms.
For this case,
we have divided the $4\pi$ total solid angle into $842$ almost
equal solid angles, using the icosahedral covering of 
Ref.\,\onlinecite{HARDIN}.  We have chosen this number since it
corresponds as close as possible to the typical number of
surface atoms in the cluster (see Table\,\ref{tab:NUMBER}). 
At each temperature over $1000$ instantaneous configurations, sampled at equal intervals throughout
the simulated time of $43$ ns, have been included in our average. 
We show the resulting average shapes in Fig.\,\ref{fig:AVERAGE}.
We present results for the following temperatures:
$400$ K, representing the low temperature configuration in
which thermal fluctuations are negligible; $600$ K where one
starts to notice small changes in the surface; $900$ K where
substantial softening of the surface orientational order
parameter $Q_6$ is observed;  $1060$ K, just below the
melting $T_{\rm m}\simeq 1075$ K; and $T=1100$ K just above $T_{\rm m}$.

In the top row of Fig.\,\ref{fig:AVERAGE} we show pictures
constructed similarly to that in Fig.\,\ref{fig:DIST}(c) of
Section\,\ref{sec:AVERAGE}, which showed the average shape of
the liquid cluster at $T=1200$ K.
The small spheres represent the average position of the 
surface within the given solid angle.  Additionally, we
have now shaded these spheres according to the value of the
maximal local surface curvature, as we did earlier in Fig.\,\ref{fig:IH}(a)
for an instantaneous configuration at $T=200$ K; the darker
the gray scale, the larger the curvature.  This method of shading
is used to highlight any edges and facets that are on the cluster surface.
The view point for these pictures is taken at infinity, so as
to  show a full hemisphere of solid angle.

In the bottom row of Fig.\,\ref{fig:AVERAGE} we show the
corresponding average shapes using a smooth $3$D contour plot
with overhead lighting.
The view point for these bottom row pictures is now taken
to be a finite distance from the cluster, in order to highlight
the straight edges and $5$-fold symmetry about the vertices.

The  pictures in Fig.\,\ref{fig:AVERAGE} illustrate the following scenario 
as the cluster is heated.  At low temperatures the cluster is almost
fully faceted, with flat facets meeting at sharp edges and vertices.
By $900$ K the facets have shrunken slightly in size and the edges and vertices
have noticeably rounded.  At $1060$ K, just below melting, the facets have
shrunken to almost negligible size, and the cluster is almost spherical.
Above melting, the cluster is essentially a perfect sphere.

\protect

\begin{figure}
\epsfxsize=8.6truecm
\epsfbox{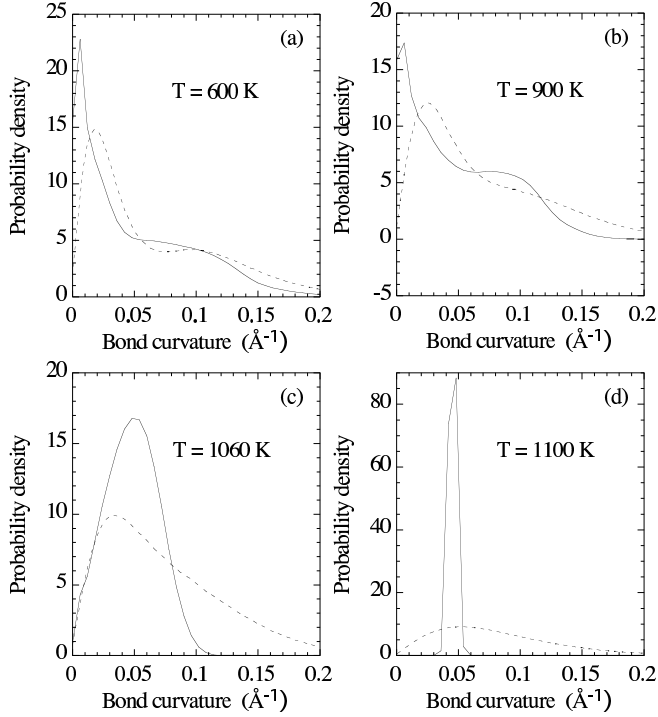}
\caption{Histograms of bond curvature $c_b$ for the average cluster shape (solid lines)
and the instantaneous cluster configurations (dashed lines) at (a) $T=600$K,
(b) $T=900$ K, (c) $T=1060$ K, and (d) $T=1100$ K.  The cluster size is $N=2624$ atoms.}
\label{fig:BONDHIST}
\end{figure}

\begin{figure}
\epsfxsize=8.6truecm
\epsfbox{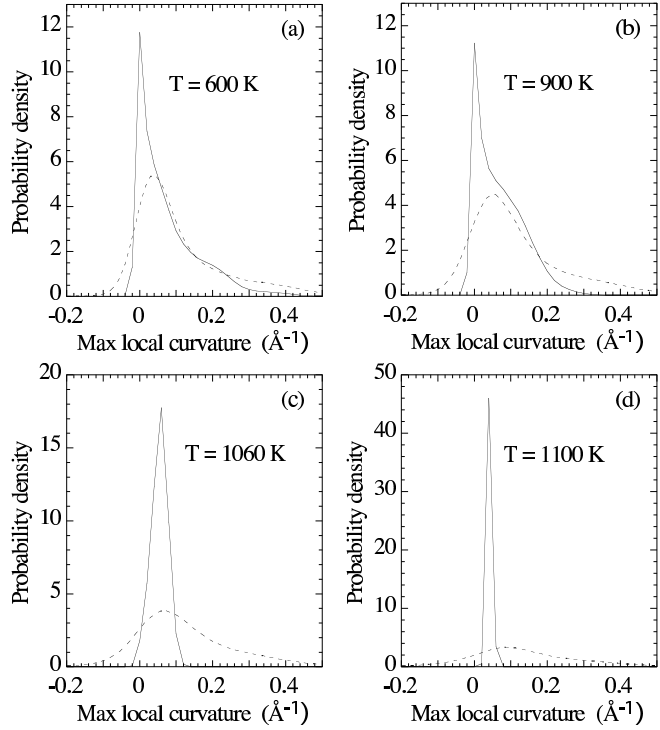}
\caption{Histograms of maximal surface curvature $\kappa_{\rm M}$ of the average cluster
shape (solid lines)
and the instantaneous cluster configurations (dashed lines) at (a) $T=600$K,
(b) $T=900$ K, (c) $T=1060$ K, and (d) $T=1100$ K.  The cluster size is $N=2624$ atoms. }
\label{fig:LOCALHIST}
\end{figure}

As a way to quantify the cluster shapes we  have computed the bond curvatures $c_b$
and the maximal local surface curvatures $\kappa_{\rm M}$, as defined in Section\,\ref{sec:ROUGH}.
In Figs.\,\ref{fig:BONDHIST}(a)-(d) we show histograms of bond 
curvature $c_b$ for
the four temperatures $600$, $900$, $1060$ and $1100$ K.
The solid curves show the histograms of bond curvatures as computed
over the surface bonds of the {\em average} cluster shape shown in
Fig.\,\ref{fig:AVERAGE}.  In contrast, the dashed curves show the  
histograms  of bond curvatures computed for an {\em instantaneous} cluster
configuration, and then averaged over the $1000$ instantaneous configurations
saved in our simulated time of $43$ ns.
Note that for the histograms for the average shape, where since we are
dealing with only one average configuration we have
relatively few points in our histogram, we have smoothed
our data using a Gaussian smoothing function with a width of $4$ bins.
The bin size here is $0.006$ \AA$^{-1}$.
In Figs.\,\ref{fig:LOCALHIST}(a)-(d) we show the analogous histograms for
the maximal local curvature $\kappa_{\rm M}$.  
The bin size here is $0.02$ \AA$^{-1}$.
Note that $\kappa_{\rm M}$ can be negative, corresponding to a region where
the surface is locally concave; an example of when this can happen is
near a vertex which is missing its top most atom.

Both Figs.\,\ref{fig:BONDHIST} and \ref{fig:LOCALHIST} 
illustrate the same scenario. Consider first the histograms of the
average cluster shapes.  At low temperatures,
the histograms show a strong peak at zero, representing the low
curvatures of the large flat facets.  The histograms also show either
a second peak or plateau at higher curvature, with a long high 
curvature tail, representing the large curvatures at edges and vertices.
We can compare these results against those in Fig.\,\ref{fig:IDEAL}
for the ideal Ih structure.  In the liquid above $T_{\rm m}\simeq 1075$ K,
the histograms have a single sharp peak at finite curvature,
representing the uniform curvature of the spherical liquid cluster.
Just below melting, at $T=1060$ K, the histograms similarly show a 
single peak near that of the liquid, only noticeably broader than for the liquid;
this indicates the shrinkage of the flat facets to negligible size and a
rounded cluster that is not yet a perfect sphere.

Comparing the histograms for the average vs. the instantaneous shapes,
the latter are in general broader, most especially for the liquid cluster.
This demonstrates the presence of strong thermal shape fluctuations about the
average shape.  Particularly interesting are the histograms for
$1060$ K, just below $T_{\rm m}$, in Figs.\,\ref{fig:BONDHIST}(c) and \ref{fig:LOCALHIST}(c), and for $1100$ K, just above $T_{\rm m}$, in
Figs.\,\ref{fig:BONDHIST}(d) and \ref{fig:LOCALHIST}(d).
The histograms for the average shape are symmetric Gaussian like peaks
about an average curvature, corresponding to a spherical or nearly spherical
cluster.  
The histograms for the instantaneous configurations, however, are skewed in
shape.  They have a low curvature peak and a broad high curvature tail,
somewhat similar to what is seen at lower temperatures.
This suggests that the instantaneous configurations can still develop small 
local facets on the surface. A similar observation has previously been made
by Lewis {\em et al.} for smaller clusters \cite{BARRAT}.
Fluctuations of the edges and vertices of these local facets lead to
an effective diffusion of the facet upon the cluster surface; averaging over these
fluctuations results in a smoothing out of the facets to negligible size when one
considers the {\it average}, rather than the instantaneous, cluster shape.  We
have seen evidence for this scenario by visual inspection of instantaneous cluster
configurations. For example, in the instantaneous liquid cluster configuration
shown in Fig.\,\ref{fig:DIST}(b) one clearly sees a flat edge along the bottom.  

\begin{figure}
\epsfxsize=8.6truecm
\epsfbox{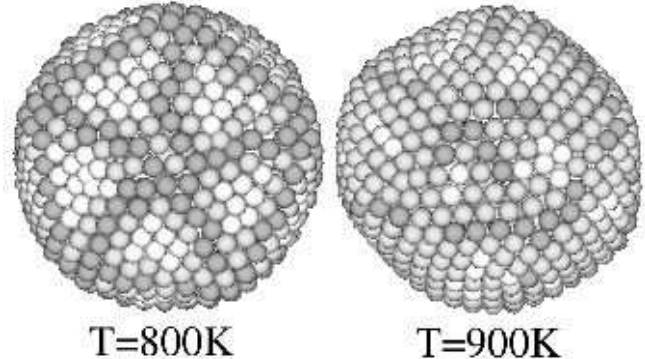}
\caption{Average cluster shapes for an $N=1409$ atom cluster at temperatures 
$800$ K and $900$ K, where $T_m\simeq 925$ K.}
\label{fig:AVERAGE1409}
\end{figure}

\begin{figure}
\epsfxsize=8.6truecm
\epsfbox{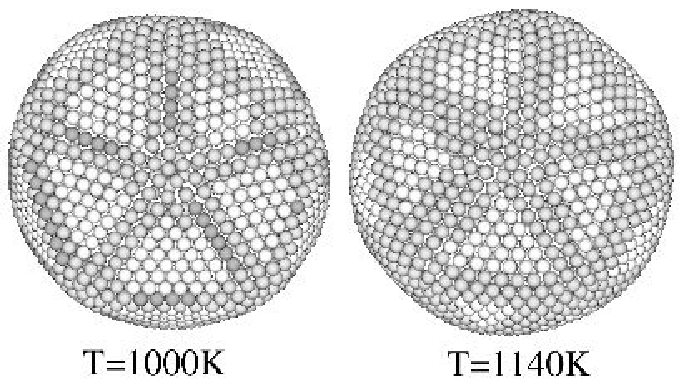}
\caption{Average cluster shapes for an $N=5082$ atom cluster at temperatures 
$1000$ K and $1140$ K, where $T_m\simeq 1150$ K.}
\label{fig:AVERAGE5082}
\end{figure}

For comparison with other sizes, we show in Fig.\,\ref{fig:AVERAGE1409}
average cluster shapes for our $N=1409$ atom cluster at temperatures $800$ K
and $900$ K, where $T_{\rm m}\simeq 925$ K.  In Fig.\,\ref{fig:AVERAGE5082}
we show average shapes for our magic number\cite{missingatom} $N=5082$
atom cluster at temperatures $1000$ K and $1140$ K, where now $T_{\rm m}\simeq 1150$ K.
The gray scale in these figures is the same as used in Fig.\,\ref{fig:AVERAGE}.
Again we see facets shrinking, and the cluster becoming more spherical, as $T_{\rm m}$
is approached.

\subsection{Diffusion of atoms}

In this section we present further evidence that the physical mechanism behind
the surface softening is indeed the diffusion of atoms on the vertices and edges of the cluster.
We will consider in this section only the cluster of $N=2624$ atoms.

\begin{figure}
\epsfxsize=8.6truecm
\epsfbox{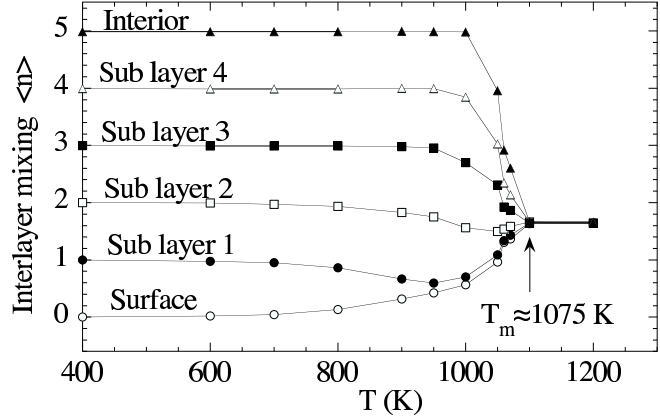}
\caption{Interlayer mixing parameter $\langle n\rangle$ vs. $T$, for 
atoms initially on the surface, in the first sub layer, \ldots, and in the interior.
The cluster size is $N=2624$ atoms.}
\label{fig:INTER}
\end{figure}

We start by first considering the {\em inter}-layer mixing of atoms in the cluster,
defining an inter-layer mixing parameter $\langle n\rangle$ as follows.  At each 
temperature we label the atoms in the initial configuration
by an integer $n^{\prime}=0,1,2,\ldots,5$ 
according to whether the atom is on the surface, in the first sub layer,
second sub layer,\ldots, or interior
of the cluster. At the end of the simulation for that temperature, we 
assign a new integer $n$ to each atom, according to which layer the
atom is now in. In Fig.\,\ref{fig:INTER} we plot $\langle n\rangle$ 
vs. $T$, where $n$ is averaged separately over each group of atoms 
indexed by their initial layer number $n^{\prime}$.  When $\langle n\rangle$
differs noticeably from the initial $n^\prime$, it indicates
significant inter-layer mixing of the atoms from layer $n^\prime$
into other layers.  From  Fig.\,\ref{fig:INTER} we see that 
noticeable inter-layer mixing takes place between the surface and the
first sub layer as low as $700$ K; these two layers are almost evenly mixed 
by $950$ K, more than $100$ K below the melting $T_{\rm m}\simeq 1075$ K.
As $T_{\rm m}$ is approached, additional 
layers start to mix together.  At $T_{\rm m}$ and above, all layers 
are evenly mixed during the course of the simulation, indicating that in
the liquid all atoms diffuse equally throughout the entire cluster.

\begin{figure}
\epsfxsize=17.2truecm
\epsfbox{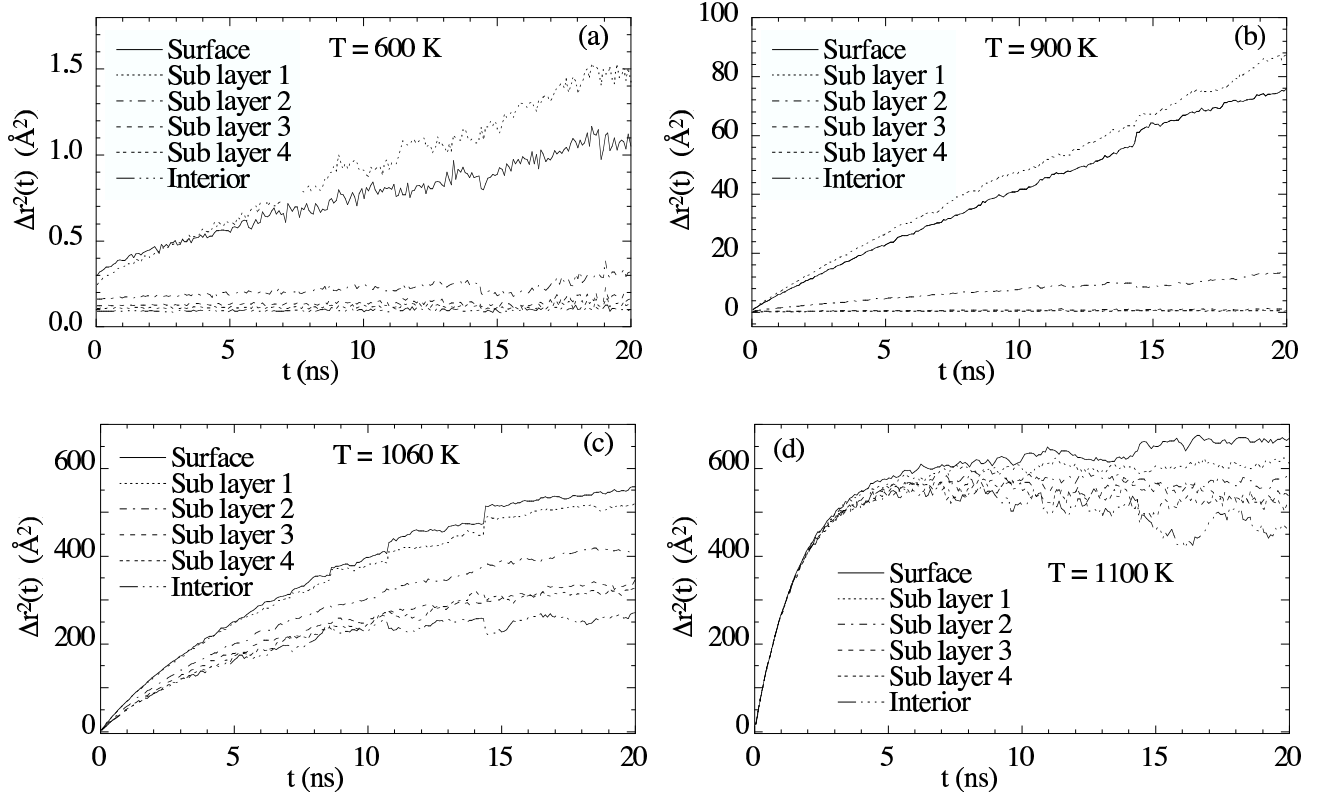}
\caption{Mean squared displacements for the $N=2624$ atom cluster averaged 
over the atoms in the surface layer, first through fourth sub layers, and interior
for (a) $600$ K, (b) $900$ K, (c) $1060$ K, and (d) $1100$ K.}
\label{fig:MSD}
\end{figure}

Next we consider the diffusion of the atoms in the cluster by computing 
the mean squared displacements, 
$\Delta r^2(t)$, defined in Eq.\,(\ref{eqn:MSD}).
We compute $\Delta r^2(t)$ separately for each layer of the cluster (and the interior)
by averaging only over the atoms that are {\em initially} in a given layer.  
In Figs.\,\ref{fig:MSD}(a)-(d) we plot our results for $\Delta r^2(t)$ vs. $t$, layer
by layer, for the four different temperatures, $600$, $900$, $1060$ and $1100$ K.
Note that since atoms in different layers can mix (see Fig.\,\ref{fig:INTER}),
the division into different layers in Fig.\,\ref{fig:MSD} contains some
ambiguity; an atom initially in the first sub layer, for example, might during
the course of the simulation wind up on the surface, however we continue to average
its motion in with that of the first sub layer.

\begin{figure}
\epsfxsize=8.6truecm
\epsfbox{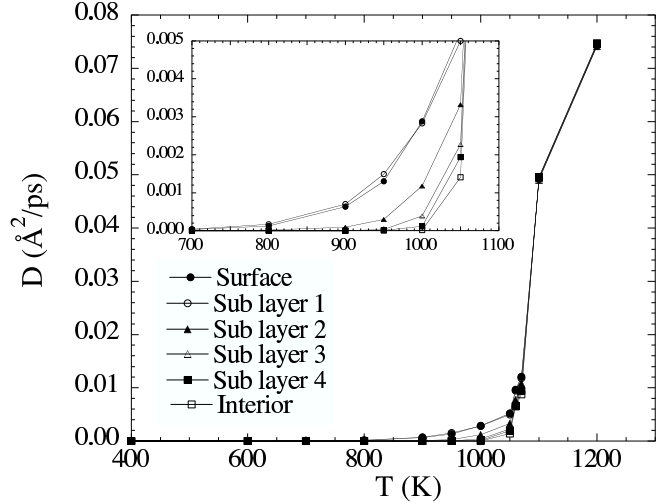}
\caption{Diffusion coefficients $D$ vs. $T$
for different layers of the $N=2624$ atom cluster. 
The inset shows an expanded range for $D$ in
a temperature range below melting, $700$ -- $1050$ K.}
\label{fig:DIFF}
\end{figure}

Several expected features are apparent in Fig.\,\ref{fig:MSD}.
In Fig.\,\ref{fig:MSD}(d) at $1100$ K, above the melting $T_{\rm  m}\simeq 1075$ K,
we see that all layers behave roughly the same, 
saturating at $\Delta r^2\sim 600$ \AA$^2$.
This is consistent with a liquid cluster of radius $\sim 21$ \AA, in which all the
atoms can diffuse throughout the entire cluster, no matter which layer they were
initially in.  
At the low temperature of $600$ K, where the average cluster shape remains almost
fully faceted, the results in Fig.\,\ref{fig:MSD}(a) show that
diffusion is almost negligible. Even for the  two
top layers, atoms on average move less than one inter-atomic spacing 
($\sim 3$ \AA)  over the observation
time of $20$ ns.  At $900$ K, where the edges and vertices of the average
cluster shape have noticeably rounded, we see in Fig.\,\ref{fig:MSD}(b) 
that diffusion in the top two
layers is significant, with atoms on average traveling a root mean square distance 
equal to several inter-atomic spacings.  The second sub layer also shows a
noticeable diffusion but all more inward atoms diffuse negligibly.
At $1060$ K, just below the melting $T_{\rm m}\simeq 1075$ K, we see
that all atoms are diffusing a significant amount throughout the cluster,
with the top two layers almost reaching the long time saturation value  
$\sim 600$ \AA$^2$ found in the liquid.

In Fig.\,\ref{fig:DIFF} we plot the diffusion constant $D$ vs. $T$ for each of the cluster
layers, obtained by fitting to the early time linear part of the curves in Fig.\,\ref{fig:MSD}.
If we fit our diffusion constant for the surface layer to the simple form
$D=D_0{\rm exp}(-E_{\rm A}/k_{\rm B}T)$ to extract the activation energy,
$E_{\rm A}=-d(\ln D)/d(1/k_{\rm B}T)$, we find the values of $E_{\rm A}=0.21$ eV
at low temperatures, $\sim 500$ K, where the cluster is fully faceted.  At high
temperature, $\sim 1200$ K, in the liquid, we find $E_{\rm A}=0.35$ eV.  Note
that the first value corresponds to surface diffusion, 
while the second value corresponds to {\em bulk} diffusion in the
liquid (sine once the cluster has melted, atoms initially on the surface easily
diffuse into the bulk).  To compare with previous simulations, 
Boisvert et al.\cite{BOISVERT} did a first principles calculation for the 
gold $\{111\}$
surface at low temperatures and found $E_{\rm A}=0.22\pm 0.03$ eV, in good agreement
with our value.  Chushak and Bartell\cite{BARTELL} reported the 
value of $E_{\rm A}=0.25$ eV
using the EAM model for a liquid gold nanocluster.  Considering the tendency of the
EAM model to systematically give lower energy values 
(as pointed out in Ref.\,\onlinecite{KELLOGG}),
our result for the liquid is in reasonable agreement.

A seeming paradox concerning our diffusion results of Fig.\,\ref{fig:MSD} is 
that at low temperatures the diffusion of atoms in the first sub layer appears to 
be greater than that for atoms on the surface. This can be explained by noting 
that the $\Delta r^2$ in Fig.\,\ref{fig:MSD} represent an {\em average} over {\em all} 
the atoms in a given layer.  As we will show below,  at low temperatures,
atoms along the edges and on the vertices
of a given layer are more mobile than a typical atom in that layer. Since the fraction of such 
edge and vertex atoms is larger in the first sub surface layer than on the surface, atoms in this layer 
have a larger average mobility. 
When the temperature increases to $1060$ K, most of the atoms in the two layers are now 
diffusing, and the average mean squared displacements $\Delta r^2$ of the two 
layers become roughly equal.

A more serious issue is how to reconcile our results of Fig.\,\ref{fig:MSD}, 
showing noticeable surface diffusion below $T_{\rm m}$, with our claim
that the surface $\{111\}$ facets remain ordered and do not premelt below $T_{\rm m}$,
as indicated from the finite values of the bond orientational order parameters
shown in Fig.\,\ref{fig:BOP}(b).  One possibility is that the surface layer
does in fact melt at a well defined temperature below $T_{\rm  m}$, but that orientational
order in the liquid surface is maintained due to the presence of an effective
periodic substrate formed by the ordered sub layers below the surface.
However we do not believe that this is the case.  Even if orientational order
in a liquid surface were preserved by the presence of the ordered sub layers,
one would still expect to see some kink or other feature in the bond orientational
order parameters at the surface melting transition.  In contrast, we find in Fig.\,\ref{fig:BOP}b
that the bond order parameters go smoothly, though near $T_{\rm m}$ steeply,
to zero.  Instead of the above scenario, we believe that the surface diffusion that we 
observe below $T_{\rm m}$ is due not to atoms on the $\{111\}$
facets, but rather due to the atoms along the vertices and edges of the surface.
As temperature increases, the facets shrink in size, the edges get rounder and broader, 
and the effective number 
of such diffusing atoms increases.  Just below $T_{\rm m}$ the facets have shrunk to
almost negligible size, the bond order parameters have decreased to a corresponding
small but finite value, and most of the atoms on the surface are now diffusing.

\begin{figure}
\epsfxsize=8.6truecm
\epsfbox{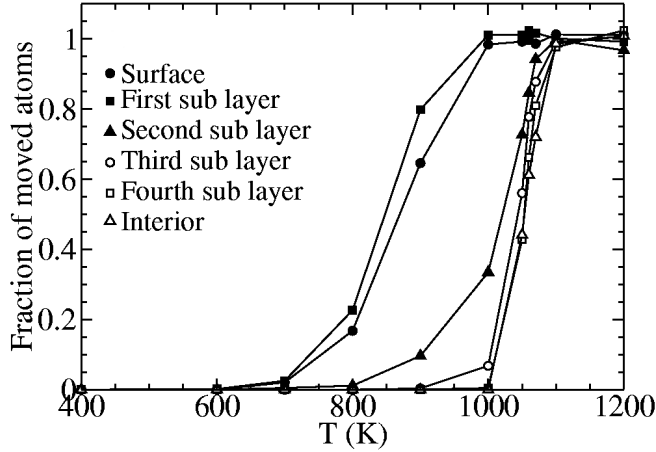}
\caption{Fraction of atoms per layer that have moved more than $8$ {\AA}
from their initial position, in $20$ ns,  for the $N=2624$ atom cluster.}
\label{fig:2624}
\end{figure}

To estimate the number of atoms in each layer that are diffusing, we use the
following criterion.  We compute the number of atoms in each layer that
have moved a distance  of more than $8$ {\AA}  within $20$ ns of simulated time.
The cutoff of $8$ {\AA}  is chosen since it is the distance between the third and fourth
peak of the pair correlation function, thus representing a distance roughly
between the third and fourth nearest neighbor; we assume that an atom which
can move this far is in fact diffusing, rather than just undergoing thermal
motion about a fixed average position.  We find our results to be qualitatively
insensitive to choosing a smaller cutoff length of $6$ {\AA} (the distance between
the second and third peaks of the pair correlation function).  If Fig.\,\ref{fig:2624}
we plot the fraction of such ``moved" atoms vs. temperature, for the surface,
sub layers, and interior of the cluster.  We see that only the surface and the first
sub layer have a significant fraction of moved atoms below melting.  This fraction steadily
increases with temperature, and approaches unity at $T \sim 1000$ K, just
below $T_{\rm m}$.  We interpret the unmoved fraction as those atoms
on the ordered $\{111\}$ facets, which shrink in size as $T$ approaches
$T_{\rm m}$.  Close enough to $T_{\rm m}$, when the facets
become so small that they are only a few atoms across, it becomes easy for
atoms on or near the edge of a facet to exchange with mobile atoms in the
surrounding ``liquid" of edge atoms; hence even such facet atoms can
ultimately diffuse throughout the cluster, and  the fraction of moved
atoms can approach unity below $T_{\rm m}$.  Indeed the concept of
liquid vs. solid become somewhat ambiguous when referring to such small
surface areas as the $\{111\}$ facets just below $T_{\rm m}$.

Combining all the above we infer the following scenario for diffusion at low temperatures:
only atoms on the surface and in the first sub layer show any noticeable diffusion well
below $T_{\rm m}$.
The atoms in these two layers that diffuse are the same atoms which mix between
the two layers (see Fig.\,\ref{fig:INTER}), and these are the atoms along the
edges and vertices of each layer.  The atoms from the
first sub layer diffuse by migrating first to the surface, and then diffusing upon
the surface, until mixing back into the first sub layer.  As the temperature increases
to $T_{\rm m}$, the facets shrink and the number of diffusing edge atoms increases,
until all surface atoms are diffusing just below $T_{\rm m}$.

\begin{widetext}
\begin{figure}
\epsfxsize=17.2truecm
\epsfbox{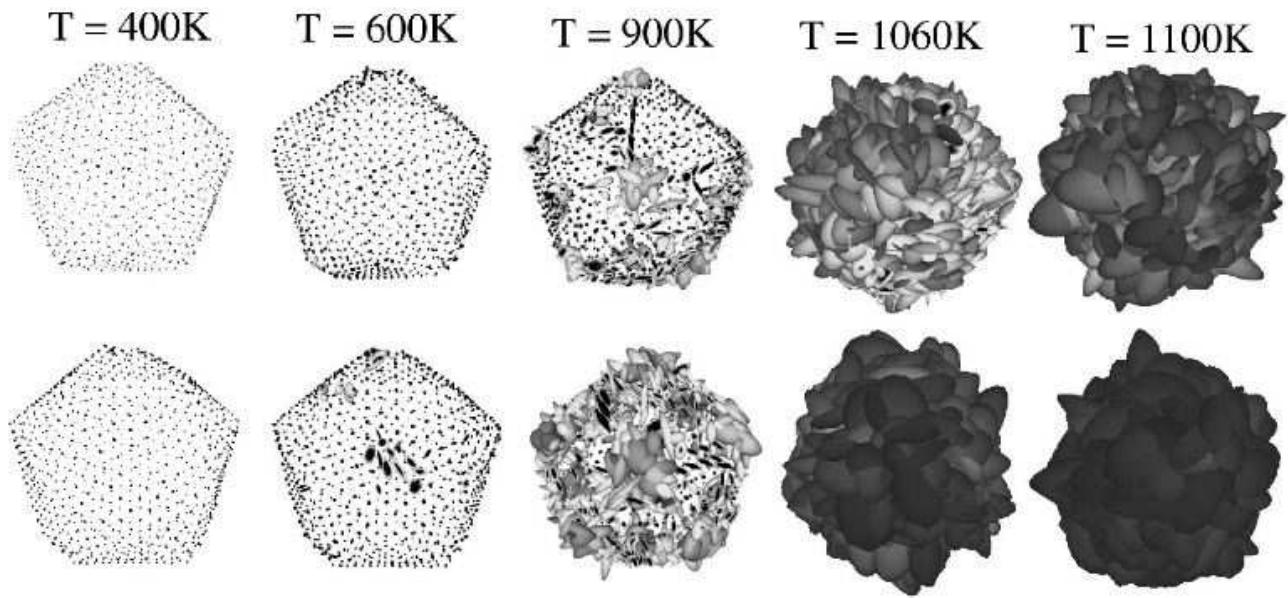}
\caption{Ellipsoids of displacement
at $400$, $600$, $900$, $1060$ and $1100$ K for the cluster of
$N=2624$ atoms. Each ellipsoid is
centered at the average position of the given atom, and shows the directional
distribution of root mean squared displacements.
The top row gives results obtained for a simulated time of $1.075$ ns,
while the bottom row is for $4.3$ ns.}
\label{fig:ELLIPSOID}
\end{figure}
\end{widetext}

To substantiate the above picture, we plot in Fig.\,\ref{fig:ELLIPSOID}
the displacement ellipsoids, defined in Section \ref{sec:DIFF}, for all
atoms initially on the surface of our $N=2624$ cluster.  We show
results for temperatures $400$, $600$, $900$, $1060$ and
$1100$ K, corresponding to the same temperatures for which we showed
the average cluster shape in Fig.\,\ref{fig:AVERAGE}.  In the top row
we show ellipsoids averaged over a simulated time of $1.075$ ns.  
We expect that atoms which are
diffusing, with $\Delta r^2 \sim t$,  should have their displacement 
ellipsoid roughly double in size when the time interval goes up by a factor
of four.  Hence in the bottom row of Fig.\,\ref{fig:ELLIPSOID}
we then show results for a simulated time of $4.3$ ns, i.e.
four times longer than the top row.  We observe the following.
At $400$ K there is no observable diffusion of surface atoms.  At $600$ K
we see diffusion of atoms at the vertices of the Ih cluster. At $900$ K we see
stronger diffusion at the vertices, as well as diffusion along the edges.  One
also can see several of the ellipsoids oriented normal to the surface, indicating
atoms which are mixing in with the first sub layer.  Atoms at the centers of
the facets remain without diffusion.  At $1060$ K and above most of the
atoms are clearly diffusing.

\protect
%


\section{DISCUSSION AND CONCLUSIONS}

We have carried out long time equilibrium molecular dynamics simulations
to study the behavior of gold nanoclusters cooled from the
liquid, and their subsequent melting upon reheating.  For
three different generic cluster sizes, $N=603$, $1409$, and $2624$, we
found that the cooled clusters formed a slightly asymmetric 
Mackay icosahedral (Ih) structure with a missing central atom.

Using the above clusters cooled from the melt, as well as several
other ``magic number'' Mackay icosahedra with up to $N=5082$
atoms that we constructed by hand \cite{missingatom}
at low temperature, we slowly heated these clusters up through
melting.  Measuring surface and bulk bond orientational order
parameters we find a sharp cluster melting transition at a temperature
$T_{\rm m}(N)$ that increases with cluster size, and that the
$\{111\}$ facets on the surface do not premelt but remain
ordered up until $T_{\rm m}$.  The surface bond parameters however
decrease from their perfect Ih values significantly 
below $T_{\rm}$, indicating a softening of the surface prior
to melting.  We find that the onset of this surface softening
appears to track the size dependent melting, occuring
roughly $~200$ K below $T_{\rm m}(N)$.

Looking at the average shape of our clusters, we see that this
surface softening corresponds to a rounding of the edges and
vertices of the cluster, with a corresponding shrinkage of the
$\{111\}$ facet area.  Just below the melting $T_{\rm m}$
the average cluster shape is nearly spherical.  As the temperature
increases towards melting, and in the liquid above $T_{\rm m}$,
instantaneous cluster configurations
can display large thermal fluctuations about this average shape.

Measuring the diffusion of atoms in the cluster, we conclude that
the mechanism for this surface softening is the onset of
diffusion of atoms at first the vertices and then the edges of 
the cluster surface, as temperature is increased.  As temperature further
increases, the mobility of these atoms increases, and more and more atoms
near the edges of the facets participate in this diffusion, until the number
of atoms remaining stationary on the facets becomes almost negligibly
small near $T_{\rm m}$.
Simultaneous with this increasing diffusion is an increase in
interlayer mixing, with surface and first sub layers mixing first,
and then deeper layers mixing in as one approaches close to $T_{\rm m}$.

A similar rounding of edges and shrinking of facets occur in the theory
of the equilibrium shape of macroscopically large crystals, where the continuum
Wulff construction \cite{WULFF} can be applied.  In this theory, the shrinkage of facets
is associated with approaching a roughening transition of the faceted
surface, and the facet length shrinks proportionally to the inverse of the
roughening correlation length \cite{JAYAPRAKASH}.  We do not believe
that this theory explains the results for our clusters.  Firstly, it is generally
believed \cite{CARNEVALI,GROSSER} that the $\{111\}$ gold surface that forms the
facets of our cluster does not have a roughening transition below the bulk
melting transition.  This is consistent with our observation that  the surface 
softening of our clusters seems to track the size dependent
cluster melting temperature rather than approaching a size independent onset
temperature as would be expected if there was a true thermodynamic roughening
transition.  Moreover, the vanishing of facets at the roughening transition
occurs within the context of the crystalline state; no diffusion of atoms need
be involved.  In our case it is clear that diffusion of atoms 
along the vertices and edges plays an important role.  
We therefore believe that the phenomena we observe in our simulated
nanoclusters are due specifically to the finite, relatively small, size of our
clusters, for which continuum approaches are not valid and one must consider
the atomistic nature of the system.  The rounding of edges and shrinkage of
facets that we observe are better attributed to a ``melting" of the cluster
edges, which then spreads out into the ordered facets as the temperature
increases towards melting.


\section{ACKNOWLEDGMENT}

This work was funded in part by DOE grant DE-FG02-89ER14017.



\begin{thebibliography}{99}
\bibitem{missingatom} Note that these values are actually one atom smaller than
    the magic number $N$ given by Eq.\,(\ref{eqn:IHN}).  This is because we construct these
    clusters with a missing central atom, in order to be consistent with the clusters
    we found from cooling, as described in Section \protect\ref{sec:missingatom}.
\end{thebibliography}
\end{document}